\documentclass[11pt]{article}
\usepackage{jheppub}
\usepackage{amsmath,amssymb,amsfonts,graphicx}

\newcommand{\be}{\begin{equation}}
\newcommand{\ee}{\end{equation}}
\newcommand{\bea}{\begin{eqnarray}}
\newcommand{\eea}{\end{eqnarray}}
\newcommand{\beas}{\begin{eqnarray*}}
\newcommand{\eeas}{\end{eqnarray*}}
\newcommand{\ba}{\begin{array}}
\newcommand{\ea}{\end{array}}

\newcommand{\cDsl}{{{\cal D}\kern-.65em /\,}}
\newcommand{\cDslsm}{{{\cal D}\kern-.5em /\,}}
\newcommand{\nabslsm}{\nabla\kern-.55em /}
\newcommand{\pasl}{\pa\kern-.55em /}
\newcommand{\psl}{p\kern-.45em /}
\newcommand{\Dsl}{D\kern-.65em /}
\newcommand{\Asl}{A\kern-.55em /}
\newcommand{\nabsl}{\nabla\kern-.65em /\kern+.2em}
\newcommand{\qsl}{q\kern-.5em /}
\newcommand{\ksl}{k\kern-.5em /}
\newcommand{\rsl}{r\kern-.5em /}

\newcommand{\cDslLCsq}{{\stackrel{\circ}{\cDsl^{\kern2pt 2}}}}

\newcommand\cc[1]{#1^{^{\kern-6pt \circ}}\kern2pt}

\newcommand{\pa}{\partial}

\newcommand{\beq}{\begin{equation}}
\newcommand{\eeq}{\end{equation}}
\newcommand{\beqn}{\begin{eqnarray}}
\newcommand{\eeqn}{\end{eqnarray}}

\def\dalemb#1#2{{\vbox{\hrule height .#2pt
\hbox{\vrule width.#2pt height#1pt \kern#1pt
\vrule width.#2pt}
\hrule height.#2pt}}}

\title{Information radiation in BCFT models of black holes}

\author[a]{Moshe Rozali}
\author[a]{James Sully}
\author[a]{Mark Van Raamsdonk}
\author[a]{\\Christopher Waddell}
\author[a]{David Wakeham}

\affiliation[\,a]{Department of Physics and Astronomy, University of British Columbia,\\
6224 Agricultural Road, Vancouver, B.C.\ V6T 1Z1, Canada.}

\emailAdd{rozali@phas.ubc.ca, jamie.sully@gmail.com, mav@phas.ubc.ca, cwaddell@phas.ubc.ca, david.a.wakeham@gmail.com}

\abstract{In this note, following \cite{GP,AE,AM}, we introduce and study various holographic systems which can describe evaporating black holes. The systems we consider are boundary conformal field theories for which the number of local degrees of freedom on the boundary ($c_{bdy}$) is large compared to the number of local degrees of freedom in the bulk CFT ($c_{bulk}$). We consider states where the boundary degrees of freedom on their own would describe an equilibrium black hole, but the coupling to the bulk CFT degrees of freedom allows this black hole to evaporate. The Page time for the black hole is controlled by the ratio $c_{bdy}/c_{bulk}$. Using both holographic calculations and direct CFT calculations, we study the evolution of the entanglement entropy for the subset of the radiation system (i.e. the bulk CFT) at a distance $d > a$ from the boundary. We find that the entanglement entropy for this subsystem increases until time $a + t_{Page}$ and then undergoes a phase transition after which the entanglement wedge of the radiation system includes the black hole interior. Remarkably, this occurs even if the radiation system is initially at the same temperature as the black hole so that the two are in thermal equilibrium. In this case, even though the black hole does not lose energy, it ``radiates'' information through interaction with the radiation system until the radiation system contains enough information to reconstruct the black hole interior.}


\begin{document}

\maketitle

\parskip=10pt

\section{Introduction}

\subsubsection*{Background}

Within the context of holographic models of quantum gravity, the formation and evaporation of black holes is a manifestly unitary process in the sense that the underlying quantum system evolves through conventional Schr\"odinger evolution with a Hermitian Hamiltonian. However, in the gravity picture, the physics of the black hole interior and the mechanism through which information about the microstate of the black hole emerges in the Hawking radiation are still not fully understood.

A crucial piece of physics to understand is the evolution of the density matrix for the black hole radiation. Hawking's original calculation \cite{Hawking:1976ra}  suggests that the entropy of this density matrix continues to increase throughout the black hole's evaporation. But unitary evolution predicts that this entropy should begin decreasing at the ``Page time'' when the black hole's (macroscopic) entropy has been reduced to half of its original value \cite{Page:1993df,Page:1993wv} and the remaining black hole becomes maximally entangled with the radiation system. The specific increasing and then decreasing behavior of the entropy of the radiation system as a function of time is known as the Page curve. Understanding how this curve comes about from the gravity picture is a key challenge.

A further mystery appeared in the work \cite{Almheiri:2012rt,Almheiri:2013hfa,Marolf:2013dba,Braunstein:2009my,Mathur:2009hf}, in which the authors argued that assuming a unitary picture of black hole evaporation leads to the conclusion that there cannot be a smooth region of spacetime behind the horizon of an evaporating black hole past the Page time. The argument was based on an apparent inconsistency between having maximal entanglement between the black hole and its early Hawking radiation after the Page time and having entanglement between field theory degrees of freedom on either side of the black hole horizon, as required by smoothness. The proposed alternative is that the old black hole develops a ``firewall'' at its horizon.

A fascinating suggestion \cite{Maldacena2013} to avoid this firewall conclusion, making use of the general idea that the connectivity of spacetime is related to quantum entanglement between underlying degrees of freedom \cite{Maldacena:2001kr, VanRaamsdonk:2010pw}, is that the entanglement between the black hole and its early radiation past the Page time is actually responsible for the existence of a smooth geometry behind the black hole horizon, in the same way that the entanglement between two CFTs in the thermofield double state gives rise to a smooth wormhole geometry connecting the two black hole exteriors.\footnote{It was suggested in \cite{VanRaamsdonk:2013sza} that this analogy could be made precise by coupling a holographic CFT to an auxiliary ``radiation'' system consisting of another copy of the holographic CFT. In this case, an initial pure-state black hole described by the first CFT would evolve to an entangled state of the two CFTs which could be dual to a two-sided black hole. In this case, the radiation system manifestly describes the region behind the horizon of the original black hole.} In this picture, the behind-the-horizon degrees of freedom are the radiation degrees of freedom, so there is no contradiction that both are entangled with outside-the-horizon modes of the black hole.

Very recently, a series of papers \cite{GP,AE,AM} have provided more detailed insight into how the black hole radiation can be seen to have an entropy described by a Page curve yet avoid the firewall paradox by the mechanism of \cite{Maldacena2013} (see also \cite{Akers:2019nfi}). The examples in these papers make use of an auxiliary radiation system coupled to a system that would otherwise describe an equilibrium black hole\footnote{See \cite{Rocha:2008fe} for an early application of this idea.}. The new insights come by making use of the quantum version \cite{Faulkner:2013ana,Engelhardt:2014gca} of the Ryu-Takayanagi formula \cite{Ryu:2006bv,Hubeny2007}, which gives the gravity interpretation of entanglement entropies for subsystems of a holographic quantum system.\footnote{For a subsystem $A$ of a holographic system, the quantum RT surface $\tilde{A}$ in the dual gravitational picture is a bulk surface which encloses a region corresponding to $A$ at the boundary of the dual spacetime and has the minimum value of the functional
\be
\label{QRT}
S_{grav}(A) = {{\rm Area} (\tilde{A}) \over 4 G} + S_{bulk}(\Sigma_A)
\ee
among extrema of this functional. Here $S_{bulk}(\Sigma_A)$ is the entanglement entropy of bulk fields in the bulk region $\Sigma_A$ enclosed by $\tilde{A}$.} Importantly, the prescription for calculating these entropies in the gravity picture requires the identification of a ``quantum extremal surface'' on which the functional (\ref{QRT}) is evaluated to calculate the entanglement entropy. A central observation of \cite{GP,AE,AM} is that during the evaporation of a black hole, the quantum extremal surface that computes the entanglement entropy of the radiation system can jump, leading to a first-order transition in the entanglement entropy that provides the necessary switch from increasing to decreasing behavior.

Further insights in \cite{GP,AE,AM} make use of the notion of the ``entanglement wedge'' of a subsystem of a holographic system, which is the portion of the full spacetime that is dual to or reconstructable from the density matrix for the subsystem, and is understood to be the bulk region enclosed by the quantum extremal surface \cite{Czech:2012bh,Wall:2012uf,Headrick:2014cta,Almheiri:2014lwa,Jafferis:2015del,Dong:2016eik,Faulkner:2017vdd}. In the examples of \cite{GP,AE,AM}, it is seen that after the transition in the quantum extremal surface, the entanglement wedge of the radiation system actually includes a portion of the black hole interior. Thus, the underlying degrees of freedom for this interior region after the transition are understood to be the degrees of freedom of the radiation system, in accord with the proposal of \cite{Maldacena2013}.

\subsubsection*{Summary and outline}

\begin{figure}
	\centering
	\includegraphics[width=95mm]{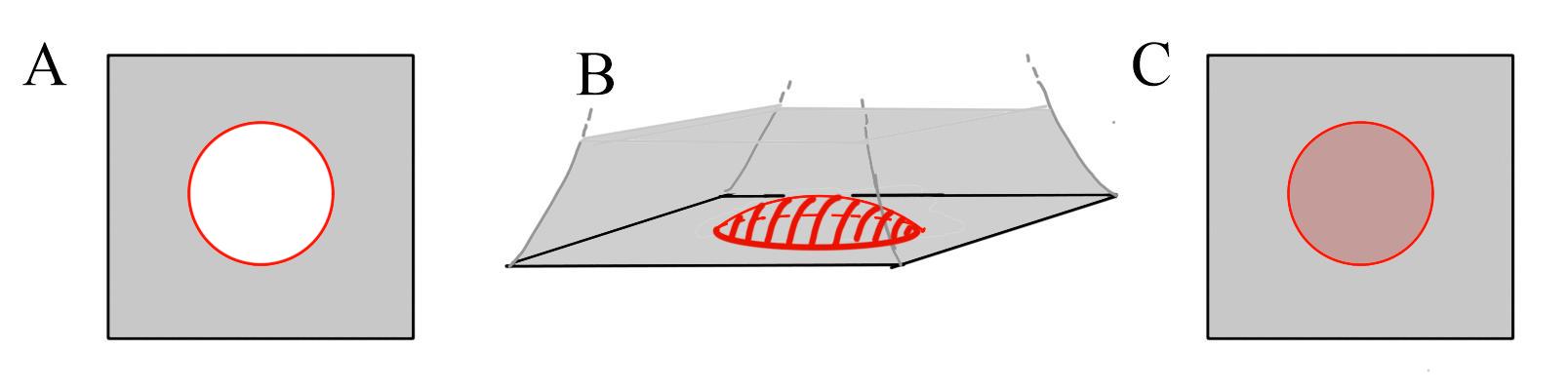}
	\caption{Basic setup. A) Our thermal system, dual to a bulk black hole, is the red boundary. It interacts with a bulk CFT which can serve as an auxiliary system to which the black hole can radiate. B) Higher-dimensional bulk picture: the red surface is a dynamical ETW brane whose tension is monotonically related to the number of local degrees of freedom in the boundary system. For large tension, this ETW brane moves close to the boundary and behaves like a Randall-Sundrum Planck brane. C) The Planck-brane picture suggests an effective lower-dimensional description where a part of the CFT in the central region is replaced with a cutoff CFT coupled to gravity, similar to the setup in \cite{AM}.}
	\label{fig:BCFT}
\end{figure}

In this paper, our first motivation is to further elucidate the observations of \cite{GP,AE,AM} by studying the evolution of black holes in a new class of models where the evolution of entanglement entropy and the entanglement wedge can be studied very explicitly through direct holographic calculations. Our models are similar to and motivated by the one in \cite{AM} in that they have a holographic description in one higher dimension than the original black hole of interest, and the full dynamics of entanglement entropy for the basic degrees of freedom is captured geometrically through the behavior of classical HRT surfaces. However, our systems are described somewhat more explicitly than the one in \cite{AM} and have an additional parameter that controls the Page time for the black hole.

Our specific construction, described in section 2, starts with a $d$-dimensional holographic system on $S^{d-1}$ in a high-energy state, or a thermofield double state with a second copy of the holographic system. These holographically describe one-sided or two-sided black holes in spacetimes that are asymptotically AdS if the theory that we start with is a CFT. The black holes are in equilibrium with their Hawking radiation, which reflects off the boundary of the spacetime. In order to have the black holes evaporate, we couple our holographic system to an auxiliary system as in \cite{VanRaamsdonk:2013sza,GP,AE,AM,Rocha:2008fe}. Our auxiliary system is a CFT in one higher dimension living on a space whose boundary is $S^{d-1}$ (or two copies of this), such that our original degrees of freedom provide boundary degrees of freedom for this higher-dimensional CFT. We can take the higher-dimensional CFT to be holographic, such that the full system is a holographic BCFT (or flows to one in the IR). We show in section 2 that the Page time for the black hole is proportional to the ratio $c_{bnd} / c_{bulk}$ of the local number of boundary degrees of freedom to the local number of degrees of freedom in the bulk CFT. In the limit where $c_{bnd}$ is large and $c_{bulk}$ is fixed, the Page time that we calculate from CFT considerations matches the Page time obtained in the gravity picture in AdS with absorbing boundary conditions \cite{Page2015}.

For our explicit calculations, we consider various states of the BCFT constructed via Euclidean path integrals, so that the dual gravity geometries can be understood explicitly. For these states, we will consider the computation of entanglement entropy for the auxiliary system, considering a spatial region defined by the points at distance greater than $a$ from the boundary system. We calculate the entanglement entropy for this system as a function of time and of the distance $a$. We perform the calculation holographically by finding the HRT surface in a dual $d+1$-dimensional gravitational system. We make use of a bottom-up holographic prescription for studying the dual BCFTs in which the CFT boundary extends into the bulk as a dynamical end-of-the-world brane whose tension is directly related to $c_{bnd}$. We also reproduce the results of these holographic calculations through direct calculations in our BCFT system, making use of standard assumptions about holographic CFTs.

\begin{figure}
\begin{center}
\includegraphics[width=.5\textwidth]{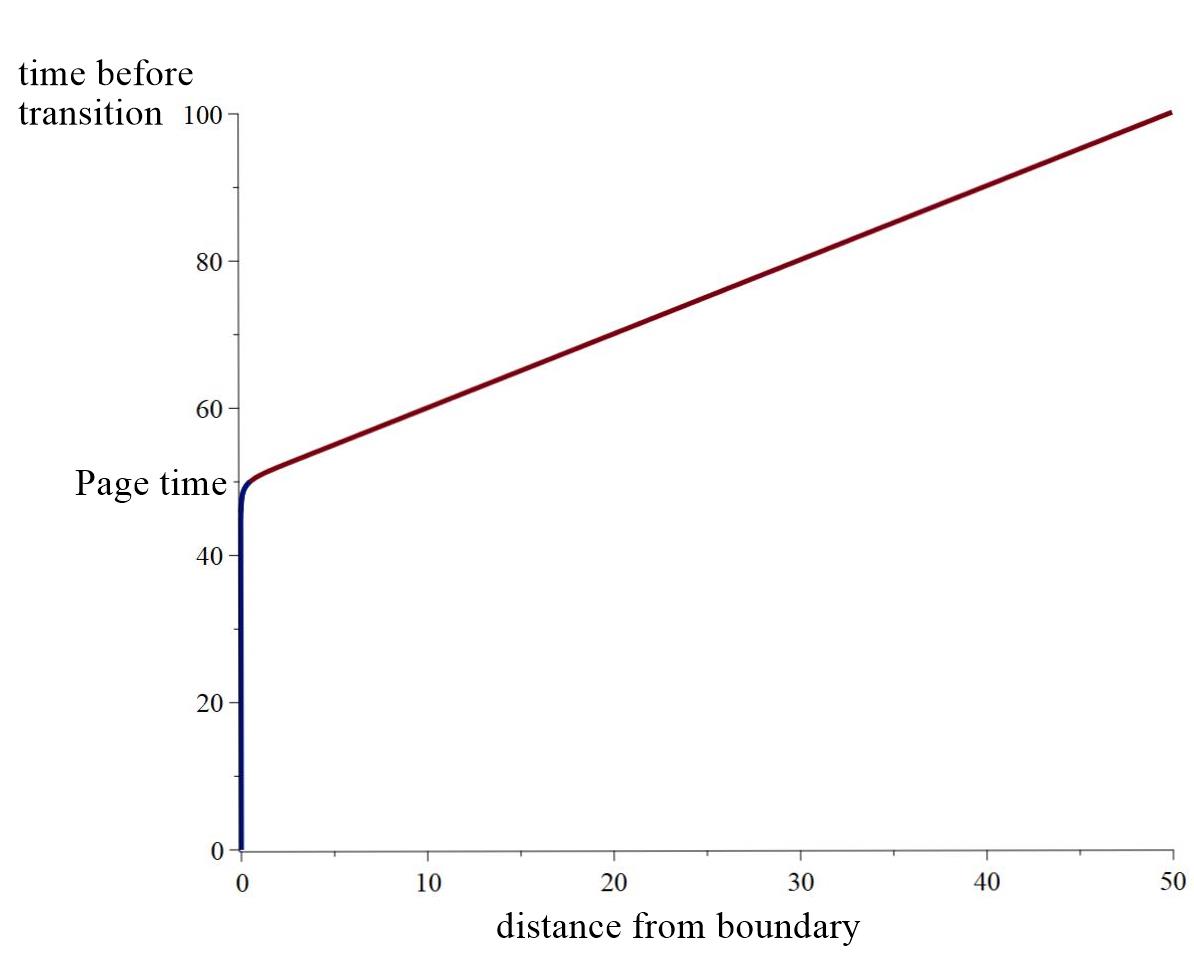}
\end{center}
\caption{Time at which the subsystem of the radiation system greater than some distance from the BCFT boundary exhibits a transition in its entanglement entropy, for the case $c_{bnd} / c_{bulk} \sim 50$. After the transition, the entanglement wedge of this subset of the radiation system includes a portion of the black hole interior. After a time equal to the Page time plus the light travel time from the boundary to our subsystem, there is enough information in the subsystem to reconstruct part of the black hole.}
\label{fig:transition}
\end{figure}

As hoped, our calculations show a first order phase transition of the entanglement entropy at the Page time after which the entropy of the radiation stops increasing; a sample result for the transition time is shown in figure (\ref{fig:transition}). In the higher-dimensional gravity picture, we find that after the transition, the entanglement wedge of the radiation system includes a portion of the black hole interior.

A new qualitative result of the present paper is that the phase transition described in the previous paragraph can occur even when the black hole is not evaporating, but simply coupled to an open radiation system which is in thermal equilibrium with the black hole. In this case, we find that while the energy density is static everywhere, the entanglement entropy for subsets of the radiation system still shows interesting dynamics, increasing with time until a phase transition after which it is constant. Again, the entanglement wedge of the radiation system includes a portion of the black hole interior after the transition. This static case is the focus of section 3.

In section 4, we consider more general states for which the initial radiation system is not in equilibrium with the black hole and the energy density is time-dependent. These more closely model evaporating black holes. Our detailed results are again in line with the expectations of \cite{GP,AE,AM} and confirm some of the qualitative predictions of \cite{AM}.

We end in section 5 with a discussion. There, we describe some directions for future work and describe further holographic constructions of evaporating black hole systems. We also point out that the transition in extremal surfaces described in this paper and in \cite{GP,AE,AM} is closely related to a similar transition \cite{Cooper:2018cmb} that can occur when looking at the entanglement entropy for subsystems of a CFT on $S^d$ in a high-energy state dual to a single-sided black hole. For the CFT states described in \cite{Cooper:2018cmb}, we can have a transition as the subsystem size is increased, after which the entanglement wedge of the subsystem includes part of the geometry behind the black hole horizon. Remarkably, in the case of 3D gravity, the CFT calculations that exhibit this transition are precisely the same CFT calculations that show the entanglement wedge transition in the present paper.

{\bf Note added:} While this manuscript was in preparation, the paper \cite{AMM2} appeared, which has some overlap with section 3 of this paper.

\section{Basic setup}

A schematic of our basic setup is shown in figure \ref{fig:BCFT}A. We imagine starting with a holographic system on $S^{d-1}$ whose high-energy states or high-temperature thermal states describe black holes in a dual gravitational picture. In these systems, the black hole is in equilibrium with its Hawking radiation, which reflects off the boundary of the spacetime.

Next, following 
\cite{VanRaamsdonk:2013sza,GP,AE,AM} we augment our holographic model with additional degrees of freedom which will serve as an auxiliary radiation system, allowing the black hole to evaporate. As in \cite{AM,AE}, our auxiliary degrees of freedom will take the form of a higher-dimensional CFT living on a space with boundary $S^{d-1}$, such that the original system now serves as a set of boundary degrees of freedom for the higher-dimensional CFT. We will denote by $c_{bulk}$ the local number of bulk CFT degrees of freedom and by $c_{bdy}$ the local number of boundary degrees of freedom. We have in mind that $c_{bdy} \gg c_{bulk} \gg 1$. This will allow the full system to be holographic, but as we show below, will give a parametrically large evaporation time.

Holographic models of this type can arise in string theory by considering branes ending on other branes. For example, we can have a stack of $n$ D3-branes in directions 0123 ending on various D5 and NS5 branes at some locations in the 3 direction \cite{Gaiotto:2008sa, Gaiotto:2008ak}. The low energy physics is ${\cal N}=4$ SYM theory on a half-space with some boundary conditions. We can have an additional $N$ D3-branes of finite extent in the 3 direction which are stretched between some of the fivebranes. Without the original $n$ D3-branes, these can give rise to a 3D CFT in the infrared. In the full setup, this 3D CFT is coupled to the ${\cal N}=4$ theory at its boundary. Here, in this setup, we have $c_{bdy} / c_{bulk} = N^2/n^2$.

\subsubsection*{Evaporation time in the CFT picture}

Now, suppose we have some initial energy $M$ in the boundary degrees of freedom such that the energy corresponds to a temperature above the Hawking-Page transition for that system. The relation between temperature, energy, and entropy is
\be
E \sim c_{bdy} R^{d-1} T^d \qquad \qquad S \sim c_{bdy} R^{d-1} T^{d-1} \; .
\ee
If this system is coupled to a higher-dimensional CFT with $c_{bulk}$ local degrees of freedom, we expect that the energy will be radiated away at a rate
\be
{d E \over dt} \sim - e c_{bulk} R^{d-1} T^{d+1}
\ee
where we are using a Boltzmann law, with emissivity $e$ that presumably depends on the nature of the coupling. The factor of $c_{bulk}$ can be understood from a weak-coupling picture where we have $c_{bulk}$ light fields that can carry away the energy.

Using these results, we have that
\be
{dT \over dt} = - \hat{e} {c_{bulk} \over c_{bdy}} T^2 \; ,
\ee
where $\hat{e}$ is defined to absorb any numerical coefficients we are ignoring.
Solving, we have
\be
T = {1 \over {1 \over T_0} + \hat{e} {c_{bulk} \over c_{bdy}}  t} \; .
\ee
The Page time is when half the (macroscopic) entropy of the black hole has been radiated. This corresponds to a temperature
\be
T_p = {1 \over 2^{1 \over {d-1}}} T_0 \; .
\ee
Ignoring factors of order 1, we find that
\be
t_{Page} \sim { c_{bdy} \over c_{bulk}}  {1 \over \hat{e}  T_0}
\ee
or
\be
\label{compare1}
t_{Page}/R \sim {1 \over c_{bulk} \hat{e}}{ c_{bdy}^{1 + {1 \over d}} \over (MR)^{1 \over d}} \; .
\ee

Since the initial energy is of order $c_{bdy}$, it is also illustrative to write $M R = x c_{bdy}$, so that
\be
t_{Page}/R \sim {  c_{bdy} \over c_{bulk} \hat{e}}{1  \over x^{1 \over d}} \; .
\ee
We see that the Page time is proportional to $c_{bdy} \over c_{bulk}$; we can make the black hole evaporation take a long time by choosing $c_{bdy} \gg c_{bulk}$.

\subsubsection*{Evaporation time for  a black hole with absorbing boundary conditions}

We can compare this to the calculation in \cite{Page2015} of Page (see also \cite{Ong:2015fha}), who considers perfectly absorbing boundary conditions for a large black hole in AdS. Using those results, one finds a Page time
\be
\label{PagePage}
t_{Page} \sim {L_{AdS}^{d+1 - {2 \over d}} \over G^{1 + {1 \over d}}}{1 \over M^{1 \over d}}
\ee
where we have omitted some numerical factors. An energy of $1/R$ in the field theory corresponds to energy $1/L_{AdS}$ on the gravity side, while field theory entropy
$c_{bdy} R^{d-1} T^{d-1}$ corresponds on the gravity side to $r_H^{d-1}/G = T^{d-1} {L^{2d -2} \over G}$ so we can relate
\be
c_{bdy} R^{d-1} = {L^{2d -2} \over G} \; .
\ee
Rewriting (\ref{PagePage}) in terms of field theory parameters, we get
\be
t_{Page}/R \sim {c_{bdy}^{1 + {1 \over d}} \over (MR)^{1 \over d}}
\ee
Comparing with the expression (\ref{compare1}) above, we see that the expressions have the same dependence on $c_{bdy}$ and $M$; to match the gravity calculation, should take $c_{bulk} e$ to be of order $1$, at least in terms of scaling with $c_{bdy}$. In order that the full system is holographic, we want to take $c_{bdy} \gg c_{bulk} \gg 1$.

\subsection{Holographic Duals of BCFTs}

In this section, we briefly review the gravitational dual description of holographic BCFTs and explain how the dual of a BCFT with large $c_{bdy} \gg c_{bulk}$ can give rise to the physics of a Planck brane whose geometry is the geometry of the black hole we are studying.

In their vacuum state, BCFTs preserve the conformal invariance of a CFT in one lower dimension. Thus, the gravity dual of a $d$-dimensional CFT with boundary in its vacuum state will generally correspond to a spacetime that is a warped product of $AdS_{d}$ with some internal space, but which has an asyptotically $AdS_{d+1}$ region with boundary geometry equal to the half space. For various supersymmetric examples, gravitational dual solutions corresponding to the vacuum state are known explicitly \cite{Chiodaroli:2011fn,Chiodaroli:2012vc}. For example, there is a family of half-supersymmetric solutions to type IIB supergravity that correspond to the vacua of ${\cal N}=4$ SYM theory living on half-space with the various boundary conditions preserving half supersymmetry (e.g. \cite{DHoker:2007zhm, DHoker:2007hhe, Aharony:2011yc,Assel:2011xz}).

In general it is difficult to work with the fully microscopic examples and to find full solutions of the ten or eleven-dimensional supergravity equations that would correspond to various BCFT states. Thus, rather than employing this top-down approach, we will consider bottom-up models of BCFT duals, introduced in \cite{Karch:2000gx,Takayanagi:2011zk,Fujita:2011fp}\footnote{Note that other bottom-up constructions for the bulk dual of a BCFT have been proposed, e.g. \cite{Astaneh:2017ghi}.}. Here, the bulk dual of a $d$-dimensional CFT with boundary is taken to be a $d+1$-dimensional gravitational theory on a space which has a dynamical boundary extending from the CFT boundary into the bulk. Just as we can consider various possibilities for the bulk gravitational effective action, we can choose various terms for the boundary effective action. We expect that for appropriate choices of the bulk and boundary effective actions, we can accurately capture the physics of various holographic CFTs.\footnote{We note that in the top-down models, there is generally not an explicit ETW brane; instead, the spacetime can ``end'' by a smooth degeneration of the internal space; the ETW brane in the bottom-up model models this higher-dimensional behavior.} In this paper, we consider the simple situation where the ETW brane couples only to the bulk metric field; its action is taken to include a boundary cosmological constant (interpreted as the brane tension) and a Gibbons-Hawking term involving the trace of the extrinsic curvature. The details of the action and equation of motion, and all the solutions that we will require in this paper may be found in \cite{Cooper:2018cmb}.

\begin{figure}
\centering
\includegraphics[width=50mm]{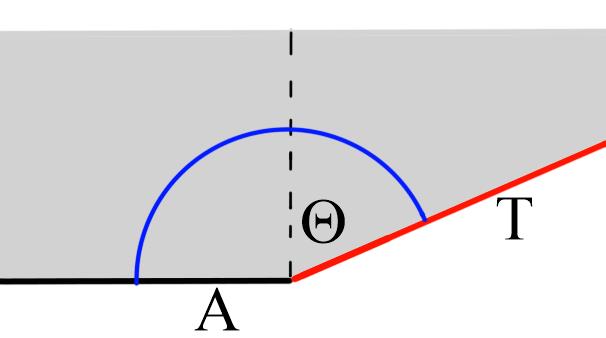}
\caption{An ETW brane with tension parameter $T$ enters the bulk at coordinate angle $\Theta$ in Fefferman-Graham coordinates. Larger $T$ gives a larger angle $\Theta$. Shown in blue is the RT surface computing the entanglement entropy of the subsystem $A$ which includes the boundary. The area to the right of the dashed line proportional to the boundary entropy.}
\label{fig:theta}
\end{figure}

The work of \cite{Takayanagi:2011zk} established a connection between the tension of the ETW brane and the boundary entropy (or higher-dimensional generalizations), which can be understood as a measure of the number of degrees of freedom associated with the boundary. One simple calculation that indicates this relation is the holographic calculation of entanglement entropy for a region of the BCFT that is the interior of a half-sphere centred on the boundary. Holographically, this is computed via the area of an extremal surface anchored to the half-sphere which extends into the bulk and ends on the ETW brane. For larger tension of the ETW brane, this brane enters the bulk at a larger coordinate angle from the vertical in Fefferman-Graham coordinates for the asymptotic region, as shown in figure \ref{fig:theta}. As a result and the area of the extremal surface becomes larger, indicating a larger boundary entropy.

In our application, we would like to consider the case where the number of local boundary degrees of freedom is large compared with the number of local bulk degrees of freedom. In this case, there is an independent way to motivate the ETW brane picture. Since we are considering the bulk CFT degrees of freedom to be much fewer than the boundary degrees of freedom, we expect that in some sense, they act as a small perturbation. Over short time scales (much shorter than the Page time), the physics of the boundary degrees of freedom is not significantly affected by the bulk CFT degrees of freedom. We can think of the  $d$-dimensional geometry of the ETW brane as the usual holographic dual of the $d-1$-dimensional boundary system in its state at a particular time. The $d+1$-dimensional system dual to the bulk CFT-degrees of freedom couples to this system, and this corresponds to adding in the bulk $d+1$-dimensional geometry coupled to the $d$-dimensional brane.
Over long time scales, the bulk CFT degrees of freedom can have a significant impact (e.g. when the black hole evaporates). Thus, over long time scales, the full geometry of the ETW brane can be affected significantly by its coupling to the bulk gravity modes, so it is important to consider the full $d+1$-dimensional system when understanding the long-time dynamics of the system.

\subsubsection*{The Randall-Sundrum Planck brane and the effective gravity picture}

As we have reviewed above, a large number of boundary degrees of freedom corresponds to a large tension for the ETW brane and in this case, the ETW brane enters the bulk at a very large angle to the AdS boundary. For the case of a single sphere-topology boundary, the resulting dual gravity solutions have ETW branes that stay close to the boundary in some sense (e.g. they correspond to a cutoff surface in a complete AdS spacetime for which light signals can propagate out to the AdS boundary and back in small proper time). In this and similar cases, the ETW brane behaves as a ``Planck brane'' in the Randall-Sundrum sense \cite{Randall:1999vf}, cutting off a portion of the asymptotic region of the geometry so that this part of the spacetime now terminates with a dynamical brane.\footnote{It is interesting that BCFTs can provide a microscopic realization of Randall-Sundrum models; this idea manifested itself in a different way in the recent work \cite{Cooper:2018cmb, Antonini2019}.} This point of view suggests a third description of the physics of our situation: from the CFT point of view, the addition of a Planck brane to a region of the bulk corresponds to cutting off the CFT in some spatial region and coupling to gravity in this region. The cutoff goes to infinity at the boundary of the region. This picture corresponds to the ``2D gravity with holographic matter'' picture of \cite{AM}. This latter picture most closely aligns with the model in \cite{AE}. The three pictures are summarized in figure \ref{fig:BCFT}. Note that it is this last picture (figure \ref{fig:BCFT}C) where the coupling between the black hole system and the radiation system is strictly at the boundary of the gravitational system.

\section{Two-dimensional models: static case}
\label{sec:2dstatic}

In this section, we will consider a very simple system that already exhibits all of the key features of the entanglement dynamics described in \cite{GP,AE,AM}. The system we consider is not an evaporating black hole, but one where the auxiliary radiation system has the same initial temperature as the black hole, so that the two systems are in equilibrium. The system we look at has a static energy density (in a particular conformal frame), but the entanglement entropy for various subsystems still evolves with time and the entanglement wedge exhibits a phase transition similar to the ones discussed in \cite{GP,AE,AM}.

\begin{figure}
\centering
\includegraphics[width=120mm]{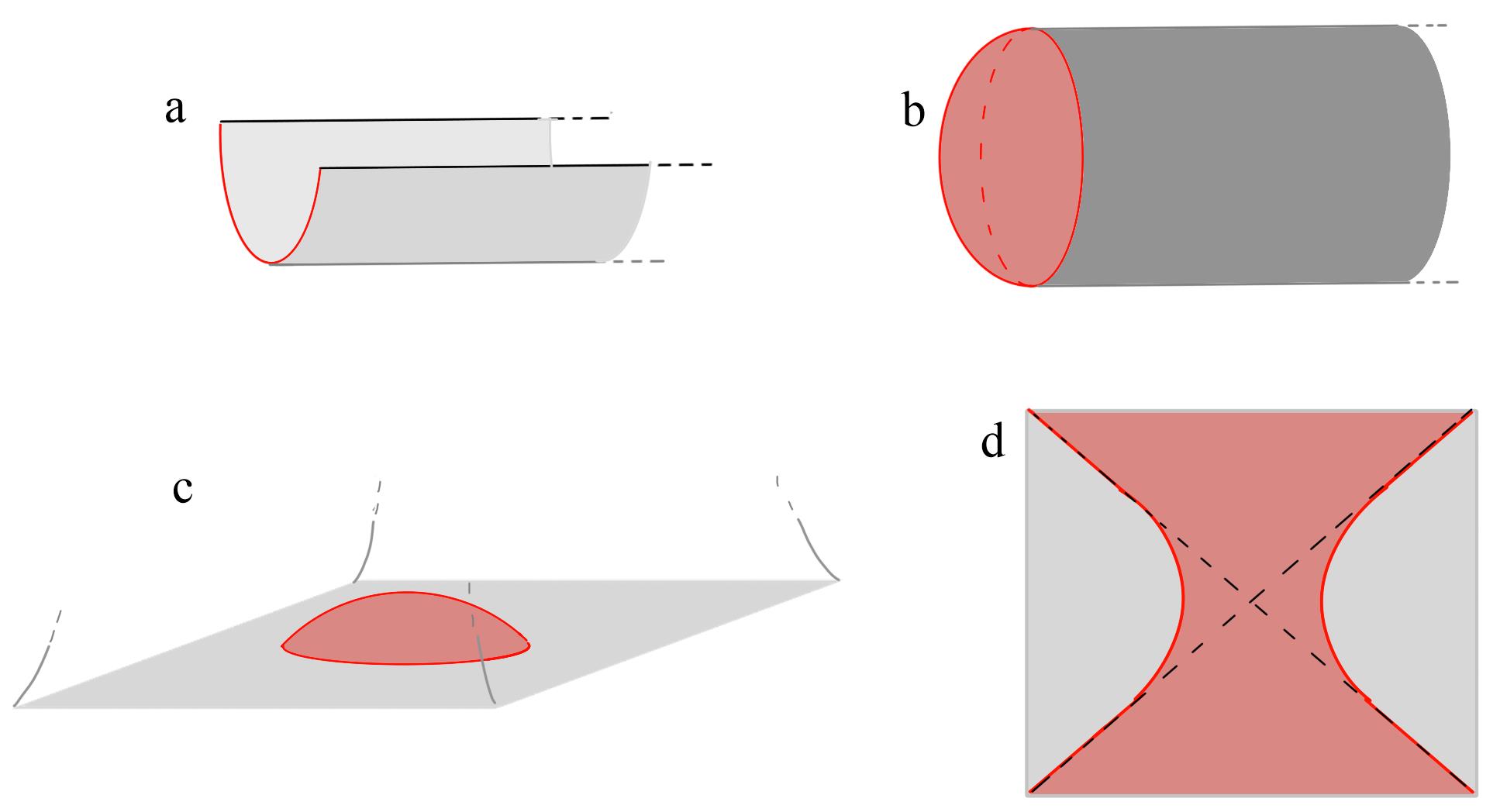}
\caption{a) BCFT path integral defining the thermofield double state of two 1+1 dimensional BCFTs. b) Euclidean geometry dual to the BCFT thermofield double. The red surface is an ETW brane. c) The same geometry represented as part of Euclidean Poincar\'e-AdS. d) Lorentzian geometry of the original state, looking perpendicular to the boundary. Dashed lines represent horizons on the ETW brane, corresponding to the horizons of the two-sided black hole represented by the boundary system.}
\label{fig:TFD}
\end{figure}

Specifically, we consider a 1+1 dimensional BCFT which is in the thermofield double state with a second copy of this system. This can be constructed via a path integral on a quarter-cylinder $y \le 0$, $0 \le \theta \le \pi$, where $\theta$ is the Euclidean time direction, and the boundary of each CFT is at $y=0$. This is shown in figure \ref{fig:TFD}a.

To understand the gravity dual, we use the bottom-up prescription where the boundary system leads to a bulk ETW brane. For 1+1 dimensional CFTs, it is convenient to define
\be
c_{bdy} = 6 \log g
\ee
where $\log g$ is the usual boundary entropy. Then, defining
\be
F = {c_{bdy} \over c_{bulk}} \; ,
\label{F}
\ee
the tension parameter $T$ (defined explicitly in \cite{Cooper:2018cmb}) for the ETW brane is related to $F$ and to the angle $\Theta$ in figure \ref{fig:theta} by
\be
\label{TFth}
T = \tanh F = \sin \Theta \; .
\ee

The dual Euclidean solution corresponding to our state is a portion of Euclidean AdS, which we may describe using metric (setting $L_{AdS}=1$)
\be
\label{AdSG}
ds^2 = (\rho^2 + 1) dy^2 + {d\rho^2 \over \rho^2 + 1} + \rho^2 d \phi^2 \; .
\ee
The specific solution we need was already constructed in \cite{Fujita:2011fp,Cooper:2018cmb}. The bulk Euclidean solution terminates on an end-of-the-world (ETW) brane with locus
\be
\label{ETW}
y(\rho) = -{\rm arcsinh} \left({\tan \Theta \over \sqrt{\rho^2+1}} \right) \; ,
\ee
where $\Theta$ is related to the brane tension and the number of boundary degrees of freedom by (\ref{TFth}). The Euclidean geometry is depicted in figure \ref{fig:TFD}b. The Lorentzian geometry dual to our state is obtained by taking the geometry of the $\phi = 0,\pi$ slice of the Euclidean solution as our initial data.

To analyze the extremal surfaces in the Lorentzian version of this geometry, it will be convenient to change coordinates to Poincar\'e coordinates, via the transformations
\be
y = \ln(r) \qquad \rho = \tan(\theta)
\ee
which bring us to spherical Poincar\'e coordinates and
\be
z = r \cos \theta \qquad x = r \sin \theta \cos \phi \qquad \tau = r \sin \theta \sin \phi \; .
\ee
which bring us to the usual Cartesian Poincar\'e coordinates in which the metric is
\be
ds^2 = {1 \over z^2}(dz^2 + dx^2 + d \tau^2) \; .
\ee
In these coordinates, the CFT boundary is at $x^2 + \tau^2 = 1$, while the ETW brane is the surface
\be
x^2 + \tau^2 + (z + \tan\Theta)^2 = \sec^2 \Theta \; ,
\ee
as shown in figure \ref{fig:TFD}c. We obtain the Lorentzian solution by analytic continuation $\tau \to i t$. This gives
\be
\label{PoincareL}
ds^2 = {1 \over z^2}(dz^2 + dx^2 - d t^2) \; ,
\ee
CFT boundary at $x^2 - t^2 = 1$, and ETW brane at
\be
\label{ETW}
x^2 - t^2 + (z + \tan\Theta)^2 = \sec^2 \Theta \; .
\ee
This is shown in figure \ref{fig:TFD}d.

\subsubsection*{Horizons on the ETW brane}

Let's now understand the causal structure of the ETW brane geometry to map out the horizons of the black hole that it contains. Consider the ETW brane in the Lorentzian picture, where it is described as the surface \ref{ETW} in the metric \ref{PoincareL}. We would like to find the future horizon for this surface, i.e. the boundary of the set of points from which it is possible to reach the right ETW brane boundary on a lightlike curve. The lightlike curves on the ETW brane satisfy
\be
x(t)^2 - t^2 + (z(t) + \tan \Theta)^2 = \sec^2 \Theta
\ee
and
\be
\left({dx \over dt}\right)^2 + \left({dz \over dt}\right)^2 = 1 \; .
\ee
We find that they are given by
\be
\label{lightlike}
x(t) = vt \pm {\sqrt{1-v^2} \over \cos \Theta} \qquad z(t) = |\sqrt{1-v^2}t \pm v \sec \Theta| - \tan \Theta \;
\ee
for $|v|<1$. The right and left boundaries of the ETW brane are described by $x = \pm \sqrt{t^2 + 1}$. The future horizons are the lightlike curves that asymptotes to this for $t \to \infty$. These are the trajectories
\be
x = \pm t \qquad z = {1 - \sin \Theta \over \cos \Theta} \; .
\ee
Thus, independent of $\Theta$, we have horizons on the ETW brane located at $x = \pm t$ and these lie at constant $z$. The black hole interior can be identified with the region $|x|<t$ or alternatively $z > {1 - \sin \Theta \over \cos \Theta}$

\subsubsection*{Extremal surfaces}

We would now like to investigate the HRT surfaces which calculate the entanglement entropy associated with the spacetime region spacelike separated from the interval $[-x_0,x_0]$ at $t=0$ (equivalently, the union of intervals $[\pm x_0, \pm \infty)$ at $t=t_0$.

In general, there are two possibilities for this HRT surface. First, we have the connected surfaces described by the semicircle
\be
t=t_0 \qquad z^2 + x^2 = x_0^2 \; .
\ee
We can also have disconnected surfaces that end on the ETW brane. We need to compare the areas to find out which one is the minimal area extremal surface that computes the entanglement entropy.

It will be somewhat simpler to perform our calculations in the Euclidean picture and then analytically continue the results to the Lorentzian case. That is, we will look at geodesics in the Euclidean geometry, evaluate their length and the length difference between the two cases, and find the phase boundary for transitions between the two surfaces. The Lorentzian version of all of these things can be obtained by analytic continuation.\footnote{We have checked that this matches with direct Lorentzian calculations.}

To find the areas, we note that the area of a geodesic semicircle of coordinate radius $R$ from the point $z=R$ of maximum $z$ to some $z_{min}$ is
\bea
\label{Areaform}
A(R,z_{\textnormal{min}}) &=& {\rm arccoth}\left({1 \over \sqrt{1 - {z_{min}^2 \over R^2}}}\right) \cr
&=& {1 \over 2} \ln \left({1 + \sqrt{1 - z_{min}^2/R^2} \over 1 - \sqrt{1 - z_{min}^2/R^2}}\right)
\eea
For $z_{min} = \epsilon$ with infinitesimal $\epsilon$, this reduces to $\ln(2R/\epsilon)$.

From this, the area of the connected extremal surface is
\be
\label{Acon}
A_{c} = \ln\left({2 x_0  \over \epsilon}\right)
\ee

\begin{figure}
\begin{center}
\includegraphics[width=.8\textwidth]{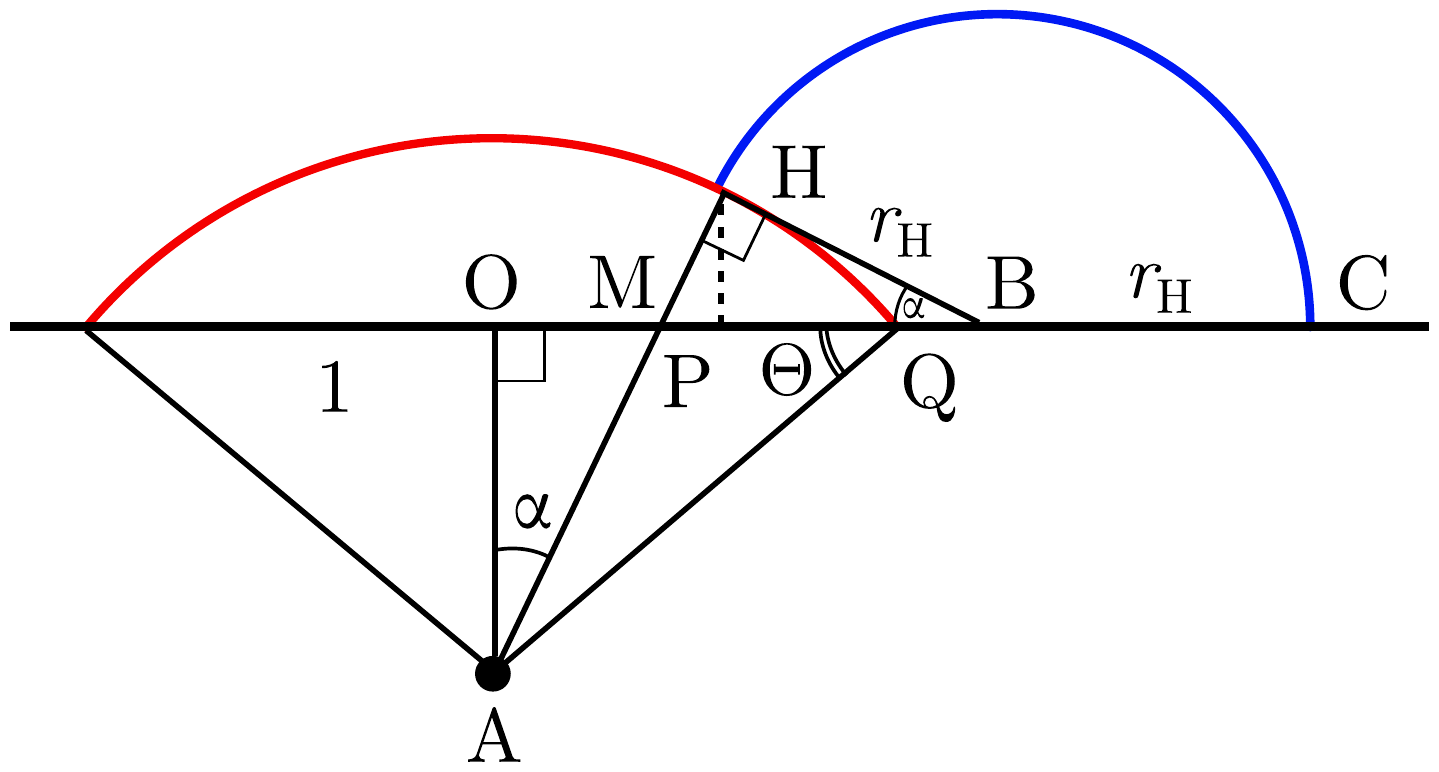}
\end{center}
\caption{Geometry of the ETW brane and half of the disconnected RT surface in the plane of the RT surface. We have $OQ=1$ and $OA=\tan \Theta$. Thus, $AQ=AH=\sec \Theta$. Also $HB \perp AH$ so $AH^2 + HB^2 = OA^2 + OB^2$. This gives $\pmb{r_H = (r^2 -1)/(2r)}$. Now $OM = OA \tan \alpha = \tan \Theta \tan \alpha$ and $AM = OA \sec \alpha = \tan \Theta \sec \alpha$. So $HM = HA - MA = \sec \Theta - \tan \Theta \sec \alpha$. Finally, $HM/HB = \tan \alpha$ gives $\pmb{ r_H = sec \Theta \cot \alpha - \tan \Theta \csc \alpha}$, while $HP = HB \sin \alpha$ gives $ \pmb{z = r_H \sin \alpha}$. The boldface equations allow us to express $z$ and $r_H$ in terms of $r$. }
\label{fig:geometry}
\end{figure}

For the disconnected surface, each part is the arc of a circle which lies at constant $\theta$, intersecting the ETW brane orthogonally and intersecting one of the the points $(\pm x_0, \tau_0)$.\footnote{In the Lorentzian picture, the disconnected RT surfaces lie at constant $x/t$ and are related by a boost to the circle arc from the point $(x=\sqrt{x_0^2-t_0^2},t=0$ to the ETW brane.} This is shown in figure \ref{fig:geometry}.

Using basic geometry (see figure \ref{fig:geometry}), we find that the extremal surface has coordinate radius
\be
r_H = {r^2 - 1 \over 2 r}
\ee
and intersects the ETW brane at $z$ coordinate
\be
z_c = {\cos \Theta \over {r^2 + 1 \over r^2 - 1} + \sin \Theta}
\ee
where $r^2 = x_0^2 + \tau_0^2$.

From (\ref{Areaform}), we find that the area of the disconnected surface (including both parts) is
\be
\label{Adisc}
A_d = \ln\left({r^2-1 \over \epsilon} {1 + \sin \Theta \over \cos \Theta}\right)
\ee
The difference in areas between the two possible extremal surfaces is
\be
A_d - A_c = \ln\left({x_0^2 + \tau_0^2-1 \over 2 x_0} {1 + \sin \Theta \over \cos \Theta}\right) \: .
\ee
From this, we see that there will be a transition when
\be
\label{EucT}
\tau_0^2 + \left(x_0 - {1-\sin \Theta \over \cos \Theta}\right)^2 = {2 \over 1 + \sin \Theta} \: .
\ee
In the Lorentzian picture, this gives the trajectory of the phase boundary as
\be
 \left(x_0 - {1-\sin \Theta \over \cos \Theta}\right)^2 = t^2 +  {2
   \over 1 + \sin \Theta} \: .
\label{LorT}
\ee
We can now map back to the original conformal frame (corresponding to figure \ref{fig:TFD}a) where the energy density is time-independent.

Using the coordinate transformations
\be
x = e^y \cos \phi \qquad \tau = e^y \sin \phi
\ee
we have that the phase boundary in Euclidean coordinates is
\be
e^F \sinh y = \cos \phi \; .
\ee
Here, $\phi$ is the Euclidean time, so in Lorentzian coordinates (where $\eta$ is the time coordinate), this phase boundary becomes
\be
e^F \sinh y = \cosh \eta \; .
\ee
Finally, if we consider an  interval $[y_0, \infty)$ (together with the equivalent interval in the other BCFT), we find that the entanglement wedge for this subsystem makes a transition to include geometry behind the black hole horizon when
\be
\eta = {\rm arccosh}( e^F \sinh y_0) \sim F + y_0
\ee
where the last relation holds for large $y_0$ and $F$. Thus, for intervals that include most of the radiation system (when $y_0$ is some small order 1 number), we see a transition at the Page time after which the black hole interior can be reconstructed from the radiation system. For large $y_0$ the time is increased by an amount which is the time taken for the radiation to reach $y_0$. The behavior of the transition time is shown in figure \ref{fig:transition}. In this frame, the entanglement entropy is constant after the transition, since each part of the disconnected extremal surface in this case is just a boosted version of the extremal surface for earlier times. Thus, the entanglement entropy increases from the initial time and then remains constant after the transition. Using the results above, the precise expression for the entropy as a function of time is\footnote{Here, we use that the cutoff surface $\rho = 1/\epsilon$ maps to the cutoff surface $z = \epsilon r$ in the Poincar\'e coordinates. We use this cutoff surface in the equations (\ref{Acon}) and (\ref{Adisc}) to calculate the entanglement entropies in the original $y$-coordinates.}
\be
S= \left\{ \ba{ll}  {c_{bulk} \over 6} \ln \left( {2 \over \epsilon} \cosh \eta \right) & \eta < {\rm arccosh}(e^F \sinh y) \cr \log g + {c_{bulk} \over 6} \ln \left( {2 \over \epsilon} \sinh y_0 \right) & \eta < {\rm arccosh}(e^F \sinh y) \ea \right. \; ,
\ee
so we have an approximately linear increase before the transition and a constant entropy afterwards.

Let's understand the physics of this phase transition in the behavior of the entanglement. We have that the energy density in both BCFTs is completely time-independent. However, the entanglement entropy for the union of regions $x > x_0$ in the two CFTs increases with time, then undergoes a first order phase transition after which it is constant. The entanglement wedge initially does not include the black hole system, but after the transition includes a portion of the interior of the black hole.

Thus, while everything is static from an energy point of view, the state is evolving in such a way that information about the black hole interior eventually becomes accessible in the auxiliary radiation system.

To understand this better, it is helpful to recall that for a free field theory in the thermofield double state, each mode in one copy of the system is purified by the corresponding mode in the other copy of the system. In our present case, we expect similarly that the boundary system is initially purified to a large extent by the other copy of the boundary system, while the bulk system is purified by the other copy of the bulk system.\footnote{Here, we are describing the situation relative to the vacuum case. Of course, there is always an infinite entanglement entropy between the boundary system of one CFT and the bulk of that CFT.} However, as we evolve forward in time, the entanglement structure evolves, and the information initially contained within the boundary system (describing our black hole initial state) leaks out into the bulk degrees of freedom, eventually leading to the transition we observe.

\subsection{Entanglement wedge after the transition}

We would now like to understand where the boundary of the entanglement wedge lies on the ETW brane after the transition.

Consider a point $(x_0, \tau_0)$ on the Euclidean transition surface (\ref{EucT}). Just after the transition to a disconnected minimal area extremal surface, the part of the surface originating at $(x_0, \tau_0)$ will end on the ETW brane at a point $(x_c, \tau_c) = \lambda (x_0, \tau_0)$. From figure \ref{fig:geometry} we see that the distance $r_c =\sqrt{x_c^2 + \tau_c^2}$ from the origin for this point will satisfy
\be
r = r_c + r_H + \sqrt{r_H^2 - z_c^2} \; .
\ee
This gives
\be
r_c = {2r \over r^2(1 + \sin \Theta) + (1 - \sin \Theta)} \; ,
\ee
so we have
\beas
\lambda &=& {r_c \over r} \cr
&=& {2 \over (x_0^2 + \tau_0^2)(1 + \sin \Theta) + (1 - \sin \Theta)} \cr
&=& {1 \over x_0 \cos \Theta + 1}
\eeas
where we have used (\ref{EucT}) in the last line. Thus, we have
\be
\label{x0xc}
x_c = {x_0 \over x_0 \cos \Theta + 1} \qquad \tau_c = {\tau_0 \over x_0 \cos \Theta + 1} \; .
\ee
Inverting these relations and plugging the resulting expressions for $x_0$ and $\tau_0$ in (\ref{EucT}), we find that the points $(x_c,\tau_c)$ lie on a curve
\be
(1 + (1-\sin \Theta)^2) x_c^2 + 2 \tan \Theta(1 - \sin \Theta) x_c + \tau_c^2 = 1 \; .
\ee
For the Lorentzian version of the problem, this becomes
\be
\label{joincurve}
(1 + (1-\sin \Theta)^2) x_c^2 + 2 \tan \Theta(1 - \sin \Theta) x_c = t_c^2 + 1 \; .
\ee
Note that $x_0 > \sqrt{t_0^2 + 1} > t_0$, so from \ref{x0xc}, we see that we will also have $x_c > t_c$. Thus, while the curve (\ref{joincurve}) crosses the horizon, the part beyond the horizon isn't relevant to us. The extremal surface always ends at a point on the brane that is outside the horizon.

Let's now calculate the proper distance to the horizon from the intersection point $(x_c,t_c,z_c)$ on the ETW brane. The ETW brane lies in the plane containing the origin and the point $(x_0,t_0)$ and extending directly inward in the $z$ direction. In this plane, the geometry is as in figure \ref{fig:geometry}, where the outermost point is at distance $r = \sqrt{x_0^2 - t_0^2}$.

This is the proper distance along the blue curve in figure \ref{fig:geometry} from $H$ to the top of the blue arc, which lies at
\be
z_{max} = \sec \Theta - \tan \Theta \; .
\ee
The distance is
\be
d = \int_{z_c}^{z_{max}} {dz \over z} \sqrt{dz^2 + dr^2}
\ee
Using
\be
r^2 + (z + \tan \theta)^2 = \sec^2 \theta \; ,
\ee
we find that the result is
\be
d = {1 \over \cos \Theta} \ln \left({r+1 \over r-1} \right) \; .
\ee
In the $y_0$ coordinates and in terms of F, this is
\be
d =  \cosh(F) \ln \left({1 + e^{-y_0} \over 1 - e^{-y_0}} \right)
\ee
We see that for large $y_0$ the location of the HRT surface intersection with the ETW brane after the transition is very close to the horizon.

Finally, we can look at the trajectory of the intersection point as a function of time after the transition. For the interval with left boundary $y_0$ in the $y$-coordinates, the initial intersection point is at
\be
x_{c} = {\sec \Theta \over 1  + {2 \over (1 + \sin \Theta)(e^{2 y_0} - 1)}}
\ee
on the curve (\ref{joincurve}) and the later trajectory follows the curve
\be
x_c^2 - t_c^2 = e^{2 y_0} (1 - x_c \cos \Theta)^2 \; .
\ee
At late times, independent of $y_0$, this approaches the point
\be
x = t = \sec \Theta = \cosh(F)
\ee
on the horizon.

\begin{figure}
\begin{center}
\includegraphics[width=.4\textwidth]{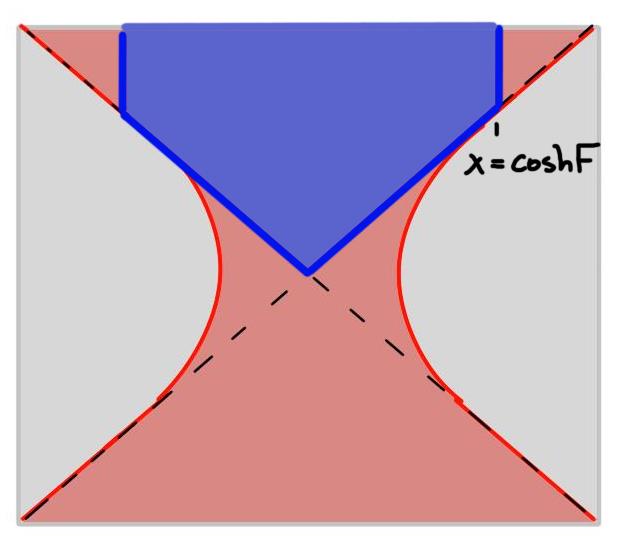}
\end{center}
\caption{The blue shaded region is the portion of the black hole interior that is included in the late-time entanglement wedge of any subsystem $|x|>a$ of the radiation system (for Poincar\'e coordinates).}
\label{fig:entwedge}
\end{figure}

The outgoing lightlike curve along the ETW brane from this point is $x=t$, while the ingoing lightlike curve along the ETW brane from this point is simply $x = \sec \Theta$ for all $t$ (using the result \ref{lightlike}). We note that the corresponding lightlike curve $x = -\sec \Theta$ on the other side of the black hole does not intersect this curve, but the ingoing lightlike curve from any closer point does intersect this curve. Thus, the points $t =  \pm x =  \sec \Theta$ are a distinguished pair of points on the horizon for which the ingoing lightlike curves barely meet at the future singularity. The late-time intersection between the entanglement wedge for the radiation system and the black hole geometry is shown in figure \ref{fig:entwedge}.

\subsection{CFT calculation}

The calculations of the previous section relied on holographic calculations of the entanglement entropy in a bottom-up holographic model where the number of boundary degrees of freedom on our BCFT is related to the tension of an ETW brane. While bottom-up models in AdS/CFT are widely studied and known to produce qualitative results that agree with those in systems that can be studied using a top-down approach, the bottom-up approach for BCFTs is less well studied, and one might thus worry whether our holographic results correctly capture the physics of genuine holographic CFTs.

In this section, we will attempt to alleviate these concerns by
reproducing our results for the entanglement entropies using direct
CFT calculations, invoking standard assumptions about the properties
of holographic CFTs.


Recall that entanglement entropy can be calculated from R\'{e}nyi entropies using the replica trick:
\[
S_A = \lim_{n\to 1} S^{(n)}_A = \lim_{n\to 1} \frac{1}{1-n}\log \mbox{Tr}[\rho^n_A].
\]
The operator $\rho^n_A$ can be related to the partition function of the $n$-fold branched cover, or \emph{replica manifold}, of the original geometry.
This, in turn, can be calculated for 2D CFTs by introducing certain \emph{twist operators} $\Phi_n$ at the entangling points of $A$ \cite{Calabrese:2004eu}.
The partition function is given by a correlator of these twists.
For $A = [z_1, z_2]$ for instance, we have
\[
\mbox{Tr}[\rho^n_A] = \langle \Phi_n(z_1)\Phi_{-n}(z_2)\rangle.
\]
In holographic theories, these correlation functions are dominated by the identity block in some channel.
A change in dominance will lead to a phase transition in entanglement entropy.
In an ordinary two-dimensional holographic CFT, this exchange causes a sudden shift from the disconnected to the connected entanglement wedge for two disjoint intervals.
In a holographic BCFT, this exchange can occur for a \emph{two-point} correlator of twists, corresponding to the entanglement entropy of a single interval.
This is analogous to the four-point result in a CFT since the two-point function in a BCFT has the same symmetries as the four-point function, and can be evaluated using the method of images.

Consider a BCFT with boundary condition $b$ on the upper half-plane
(UHP), $\{\Im(z) \geq 0\}$.
We can perform a global transformation to the complement of the disk
of radius $R$ via
\begin{equation}
  \label{eq:6}
  w = R\left(\frac{1}{z - i/2} - i\right).
\end{equation}
For simplicity, we also define $\vartheta :=w+iR$.
We then have
\begin{equation}
  \label{eq:11}
  z = \frac{R}{\vartheta}+\frac{i}{2},
\quad \Im [z(w)]
= \frac{|w|^2-R^2}{2|\vartheta|^2},
\quad w'(z) = -\frac{1}{R}\vartheta^2.
\end{equation}
Since we have performed a global transformation, the energy density vanishes:
\begin{align}
  \label{eq:2}
  \langle T(w) \rangle & = \frac{c}{12}\{z; w\} = \frac{c}{12}\frac{z'''z'- (3/2)(z'')^2}{(z')^2}  = 0.
\end{align}

Consider a two-point function of twist operators, $\Phi_n(w_1),
\Phi_{-n}(w_2)$, introducing an $n$-fold branched cover with branch
cut from $w_1$ to $w_2$.
The twists are primary by definition, so the correlation function transforms as
\begin{small}
  \begin{align}
    \langle \Phi_n(w_1)\Phi_{-n}(w_2)\rangle_{\bar{D}_G} & = |w'(z_1)
                                                         w'(z_1)|^{-d_n}\langle
                                                         \Phi_n(z_1)\Phi_{-n}(z_2)\rangle_{\text{UHP}}
    \notag\\
                                                       &  =\left|\frac{(\vartheta_1\vartheta_2)^2}{R^2}\right|^{-d_n}\langle
                                                         \Phi_n(z(w_1))\Phi_{-n}(z(w_2))\rangle_{\text{UHP}}.
  \end{align}
\end{small}
$\!\!$For holographic BCFTs, the correlator of twists on the UHP can
be evaluated \cite{bcft2}, using vacuum block dominance and an appropriate sparsity condition on the
density of states, in a similar vein to \cite{Hartman2013}.
Using this correlator and the replica trick,
the entanglement entropy of the interval $A = (-\infty, w_1] \cup
[w_2, \infty)$ is calculated by
\begin{align}
  S_A & = \lim_{n\to 1}\frac{1}{1-n}\log \langle
            \Phi_n(w_1)\Phi_{-n}(w_2)\rangle_{\overline{\text{disk}}} \notag \\
  & = \frac{c}{6}\left[2\log \left|\frac{\vartheta_1\vartheta_2}{R}\right|+ \min\left\{\frac{12}{c}g^b +
    \log
    \left|\frac{(|w_1|^2-R^2)(|w_2|^2-R^2)}{(\vartheta_1\vartheta_2\tilde{\epsilon})^2}\right|,
    \log
    \left|\frac{Rw_{12}}{\vartheta_1\vartheta_2\tilde{\epsilon}}\right|^{2}\right\}\right]
\label{global-ee}
\end{align}
where $g^b := -\log\langle 0|b\rangle$ is the boundary entropy, and $F$ is given by (\ref{F}).
We note the relations
\begin{equation}
  \label{rels-F-theta}
  e^F = \frac{1+T}{\sqrt{1-T^2}} = \frac{1+\sin\Theta}{\cos\Theta},
  \quad 1 - e^{-2F} = \frac{2\sin\Theta}{1+\sin\Theta},
\end{equation}
which we will use momentarily.
Note that a UV regulator $\epsilon$ in Poincar\'{e} coordinates becomes $\tilde{\epsilon} = |w'_{1,2}|\epsilon$ after the global transformation $z \mapsto w$ \cite{Caputa:2019avh}; the phase boundaries are unaffected.

We now specialize to the symmetric interval $A$ at some fixed time $\Im(w)
= \tau_0$, with $w_{1,2} = \pm x_0 + i \tau_0$.
Exponentiating (\ref{global-ee}), a phase transition occurs at
\begin{align}
  \label{phase-sep-1}
\left(x_0^2 - e^{-F} R\right)^2 + \tau_0^2 & = R^2(1-e^{-2F}) \\
\Longrightarrow \quad \left(x_0^2 - \frac{\cos\Theta}{1+\sin\Theta}
                                             R\right)^2 + \tau_0^2 & =
                                                                     \frac{2R\sin\Theta}{1+\sin\Theta},
\end{align}
using (\ref{rels-F-theta}).
In Lorentzian signature, $\tau_0^2 \to -t_0^2$, and we obtain
\begin{equation}
  \label{phase-sep-2}
\left(x_0^2 - \frac{\cos\Theta}{1+\sin\Theta}
                                             R\right)^2 = t_0^2 +
                                                                     \frac{2R\sin\Theta}{1+\sin\Theta}.
\end{equation}
These phase boundaries precisely match (\ref{EucT}) and (\ref{LorT})
for $R = 1$.

\subsection{Holographic replica calculation}

\begin{figure}
	\begin{center}
		\includegraphics[width=.9\textwidth]{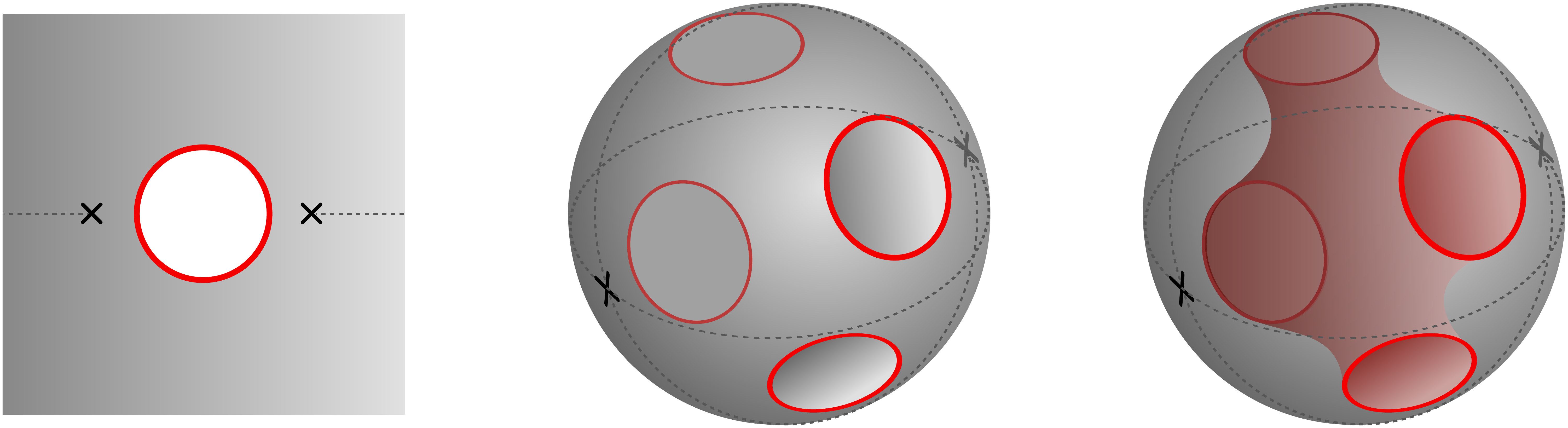}
	\end{center}
	\caption{Replica calculation of entanglement entropy.}
	\label{fig:replica}
\end{figure}

It is interesting to consider a replica version of the same calculation.\footnote{The observations of this section relating the entanglement wedge phase transition and the appearance of connected boundary saddles were directly inspired by similar observations in the JT-gravity context \cite{Almheiri:2019qdq}; related obnservations were made independently by \cite{Penington:2019kki}.}  In calculating the entanglement entropy, we want to evaluate the Renyi entropies by calculating the BCFT partition function on a replica manifold obtained by gluing $n$ copies of the Euclidean space shown in figure \ref{fig:replica} across the cut. The topology of the replica manifold is a sphere with $n$ boundaries, as shown in the second figure. Considering a larger and smaller portion of the radiation system corresponds to enlarging or shrinking the size of the boundaries relative to the size of the sphere.

Now we can consider performing this path-integral calculation holographically, using the bottom-up approach where the boundaries extend into the bulk as ETW branes. In the case of a smaller portion of the radiation system, the holes in the second picture will be small, and we will have a set of disconnected ETW branes of disk topology that ``cap off'' the boundary holes. On the other hand, as we consider a larger portion of the radiation system, the circles become large in the second picture, and we expect that the dominant saddle in the gravitational calculation will correspond to the topology shown in the picture on the right where we have a single connected ETW brane with multiple boundary components.

It seems immediately plausible that the transition to this new bulk topology is directly related to the transition of HRT surfaces in our original calculation, since the two calculations must agree.
However, it also appears at first slightly confusing: the CFT calculation correctly reproduces the disconnected bulk HRT surface from the disconnected contribution to the twist correlation function alone, while this bulk saddle is a complicated connected geometry involving both twist operators.
To align the CFT and bulk pictures, note that the same issue appears when calculating the entanglement entropy of two (or multiple) intervals in the vacuum of a 2D CFT \cite{Hartman2013}.
There, the higher Renyi entropies are also computed by a connected bulk geometry \cite{Faulkner:2013yia}, but the entanglement entropy is a sum of disconnected contributions.
This is consistent because the semi-classical Virasoro block describing the connected geometry reduces to the identity exchange in the limit $n \rightarrow 1$.
Despite the slightly different setting, the same ideas and kinematics describe the BCFT Renyi calculation \cite{bcft2}.

Thus, taking into account the second HRT surface that correctly sees the decreasing branch of entanglement entropy corresponds in the gravity version of the replica calculation to including non-trivial topologies. Had we stuck with the original topology (as we would do if treating gravity perturbatively) it seems that we would get an answer which misses the transition, and is perhaps more akin to Hawking's original calculation.

\section{2D evaporating and single sided examples}

In this section, we continue focusing on two-dimensional models, but generalize the simple example of the previous section to a case where we have a pure state of a single-sided black hole, and to cases with a dynamical energy density (as in the example of \cite{AM}) that more closely models the physics of a genuine evaporating black hole.\footnote{Of course, there are many examples that we can obtain from the previous case via local conformal transformations which would have non-trivial evolution of the energy density and may look more like an evaporating black hole. However, in this section, we focus on examples that are not conformally related to the one in the previous section.}

\subsection{Single-sided case}

\begin{figure}
\begin{center}
\includegraphics[width=.6\textwidth]{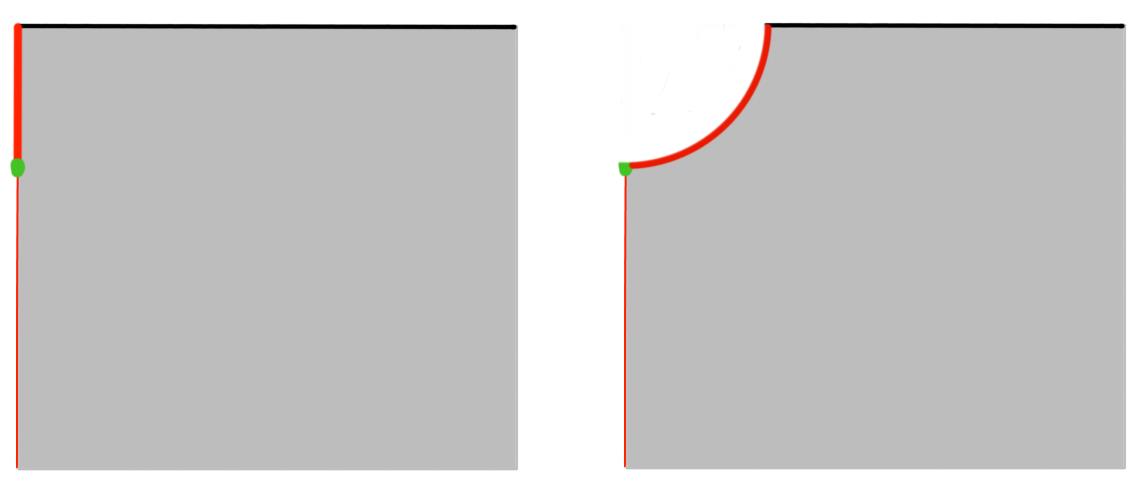}
\end{center}
\caption{BCFT models for single-sided black holes.}
\label{fig:singleside}
\end{figure}

It is straightforward to come up with BCFT examples of single-sided black holes. For example, the first picture in figure \ref{fig:singleside} shows a path-integral defining the state of a BCFT with some boundary system (fat red line) with many degrees of freedom. Here, instead of evolving the full BCFT from $\tau=- \infty$ to define the vacuum state of this system, we only evolve the boundary system from some finite past Euclidean time, as for the SYK states in \cite{Kourkoulou:2017zaj}. For prior Euclidean times, we have a different boundary condition (thin red line) that we take to be associated with a small number of boundary degrees of freedom. At the transition between these two boundaries we have an appropriate boundary condition changing operator.

This construction should place the boundary system in a high-energy state, while the bulk CFT degrees of freedom should be in a lower-energy state (through they are also affected by the change of boundary conditions in the Euclidean past). In this case, the dual gravity solution will involve ETW branes with different tensions, and a codimension-two brane associated with the boundary-condition changing operator.

It would be interesting to analyze this example in detail. For now, we point out that we can understand the physics of a very similar example using the results of the previous section. The second picture in figure (\ref{fig:singleside}) shows almost the same setup, but with a different geometry for the path-integral. This picture is similar to a $Z_2$ identification of our setup from the previous section. If we choose the lower boundary condition to correspond to a $T=0$ ETW brane in the bulk and we choose the boundary-condition changing operator appropriately (so that the equation of motion at the codimension-two brane gives a constraint that the two-types of ETW branes should meet orthogonally), then the dual geometry for this setup will be precisely a $Z_2$ identification of the bulk geometries from the previous section, with a zero-tension ETW brane at the $Z_2$ fixed point. In this case, all of our calculations and qualitative conclusions go through almost unchanged. The only significant difference is that the connected RT surface from the previous section is now replaced by its $Z_2$ identification, which ends on the $T=0$ brane.

\subsection{Dynamical case}

We can also modify our two-sided example in order to introduce time evolution of the energy density more characteristic of an evaporating black hole. We would like to have a situation where our auxiliary system starts out in a state that is closer to the vacuum state, so that the energy in the initial black hole state will radiate into this system.

\begin{figure}
\begin{center}
\includegraphics[width=.8\textwidth]{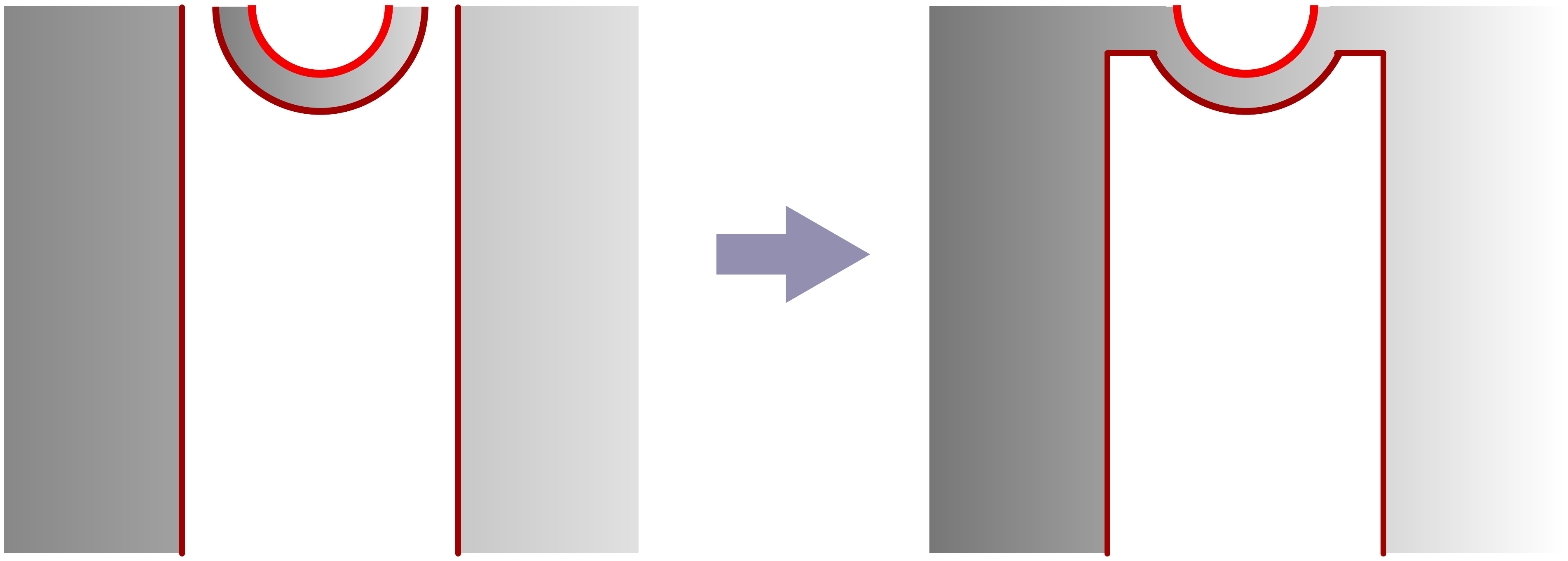}
\end{center}
\caption{2D model for an evaporating black hole.}
\label{fig:Cardy2}
\end{figure}

A simple construction (similar to that discussed in \cite{AM}) is shown in figure \ref{fig:Cardy2}. The left picture shows a state of four quantum systems. The outer systems are BCFTs with some boundary condition (denoted by a dark red boundary) that we imagine has a small boundary entropy. The path integrals shown place these systems into their vacuum state. The remaining part of the path integral constructs a thermofield double state of two systems, each of which is a BCFT living on a small interval with different boundary conditions on the two ends. The dark red boundary condition is the same as before, but the semicircular boundary (shown bright red) corresponds to a boundary system with many degrees of freedom as in the example of the previous section.

In order to make the two-sided black hole evaporate, we consider a modified system where we glue the systems together as shown on the right side of figure (\ref{fig:Cardy2}). In the final path integral, shown on the right, we are describing a state of the same system that we considered in the earlier part of this section. However, since our Euclidean path integral is in some sense a small modification of the picture on the left, we expect that far away from the black hole, the local physics of the reservoir system will be similar to the vacuum. In this case, the energy in the (bright red) boundary degrees of freedom will gradually leak out into the reservoir system. The dual gravitational picture will be that of an evaporating black hole.

In studying the dual system explicitly using the bottom-up approach, we will now have two types of branes, one with a larger tension corresponding to the blue boundary condition, and one with a smaller tension corresponding to the drak red boundary condition. The latter is what \cite{AM} refer to as the Cardy brane.  We expect that the behavior of this system should match the qualitative picture described in \cite{AM}, but now it should be possible to study everything quantitatively. Since the branes only couple to the metric and we are in three dimensions, the local geometry of the holographic dual will be that of AdS, and the dynamics of the system will be reflected in the trajectories of the ETW branes.

\subsubsection*{Phase Boundaries on the Annulus}

In order to study situations like the previous section, we can apply the methods of \cite{Shimaji:2018czt,Caputa:2019avh} who were making use of a similar Euclidean setup (without the middle boundary) to study local quenches in a holographic CFT. For any specific shape of the boundaries in (\ref{fig:Cardy2}), it is possible to map the doubled picture describing the full CFT path integral conformally to an annulus, where the circular boundary maps to the inner edge of the annulus and the other boundaries (shown in dark red) together map to the outer boundary of the annulus. We can also map the annulus to a finite cylinder, so we see that the physics will be related to the physics of the thermofield double state of a pair of CFTs on a finite interval with different boundary conditions on the two ends.

\begin{figure}
\begin{center}
\includegraphics[width=.8\textwidth]{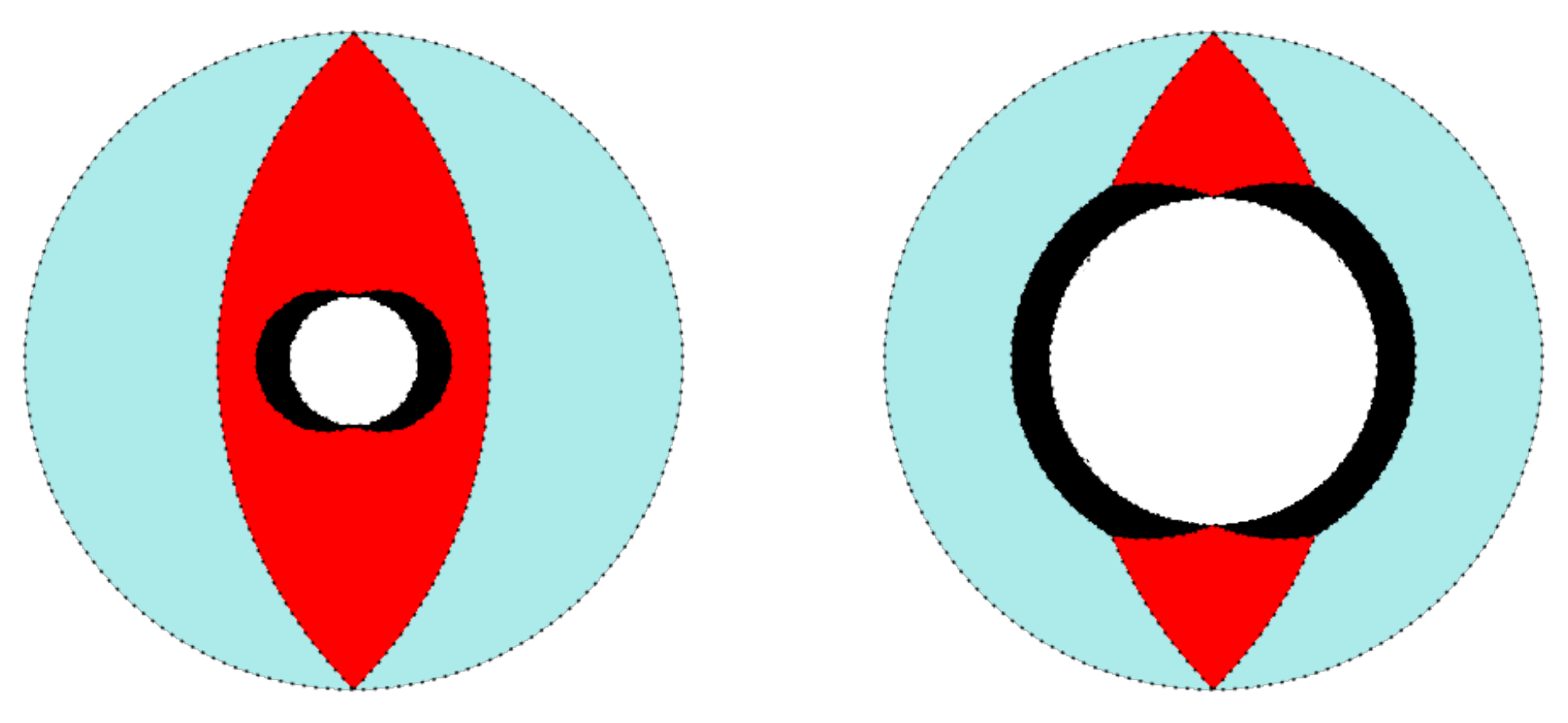}
\end{center}
\caption{Phase diagram for annulus with supercritical and subcritical $L$ respectively.}
\label{fig:annulus}
\end{figure}

We can again start with the global AdS metric (\ref{AdSG}) in which we know the ETW trajectories explicitly. Here, though, we consider a finite segment of the boundary cylinder, with a boundary condition corresponding to tension $T$ at $y=-L$ and a boundary condition corresponding to tension $T=0$ (or some other tension) at $y=0$. Changing to Poincar{\'e} coordinates as in Section \ref{sec:2dstatic}, the CFT region becomes an annulus with inner radius $R = e^{-L}$ and outer radius 1, centred at the origin. Also as in that section, the location of the ETW brane corresponding to the inner boundary is
\begin{equation}
    x^{2} + \tau^{2} + (z + R \tan \Theta)^{2} = R^{2} \sec^{2} \Theta \: , \qquad \Theta = \arcsin(T) \: ,
\end{equation}
while that corresponding to the outer boundary is
\begin{equation}
    x^{2} + \tau^{2} + z^{2} = 1 \: .
\end{equation}
For sufficiently large $L$, the two BCFT boundaries are far apart and the phase boundaries for the transition between connected and disconnected HRT surfaces are those found previously for the case of a single boundary; the phase boundary for the transition between a connected surface and a disconnected surface ending on the inner ETW brane has locus
\begin{equation}
    \Big( x - \frac{R(1-\sin \Theta)}{\cos \Theta} \Big)^{2} + \tau^{2} = \frac{2R^{2}}{1 + \sin \Theta} \: ,
\end{equation}
while that for the outer ETW brane is
\begin{equation}
    ( x + 1 )^{2} + \tau^{2} = 2  \: .
\end{equation}
(These are the phase boundaries in the region $x>0$; the $x<0$ phase boundaries are given by symmetry about $\tau = 0$.)
As $L$ is decreased to some critical value
\begin{equation}
    L_{\textnormal{c}} \equiv - \ln \Big( \frac{(-1 + \sqrt{2}) \cos \Theta}{(1 - \sin \Theta) + \sqrt{2(1 - \sin \Theta)}} \Big) \: ,
\end{equation}
the phase boundaries will osculate within the annulus at $\tau = 0$; for smaller $L$, a direct transition between disconnected HRT surfaces ending on the higher tension brane and surfaces ending on the lower tension brane can occur (see Figure \ref{fig:annulus}). The phase boundary between these disconnected phases is given by
\begin{equation}
    x^{2} + \tau^{2} = R \Big( \frac{(1 - \sin \Theta) + R \cos \Theta}{R(1 - \sin \Theta) + \cos \Theta} \Big) \equiv \ell^{2} \: .
\end{equation}

We can now map to a new conformal frame with the desired dynamical Cardy brane; the phase boundaries should simply be pushed forward using the appropriate conformal transformation, then analytically continued to Lorentzian signature.
Note \cite{Shimaji:2018czt} that, starting from Poincar\'e coordinates
\be
ds^2 = {d \eta^2 + d \zeta d \bar{\zeta} \over \eta^2}
\ee
a map $\zeta = f(w)$ corresponds to a coordinate transformation
\beas
\zeta &=& f(w) - {2 z^2 (f')^2 (\bar{f}'') \over 4 |f'|^2 + z^2 |f''|^2} \cr
\eta &=&  {4 z |f'|^3 \over 4 |f'|^2 + z^2 |f''|^2}
\eeas
in the dual asymptotically AdS geometry, which gives a metric
\be
ds^2 = {1 \over z^2} \left(dz^2 + dw d\bar{w} + z^2 ( T(w) dw^2 + \bar{T}(\bar{w}) d \bar{w}^2) + z^4 T(w) \bar{T}(\bar{w}) dw d\bar{w} \right)
\ee
where the holographic stress tensors (corresponding to the stress tensors in the CFT state) are given by
\be
T(w) = {3 (f'')^2 - 2 f' f''' \over 4 (f')^2} \qquad \qquad \bar{T}(\bar{w}) = {3 (\bar{f}'')^2 - 2 \bar{f}' \bar{f}''' \over 4 (\bar{f}')^2} \: .
\ee

\subsubsection*{Conformal mapping}

\begin{figure}
\begin{center}
\includegraphics[width=.4\textwidth]{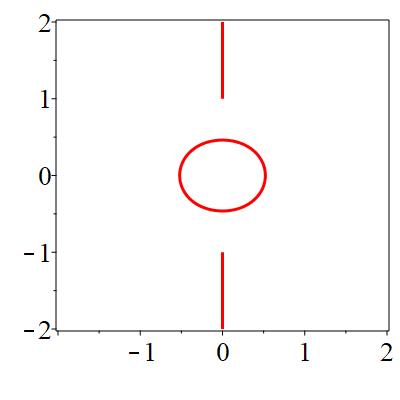}
\end{center}
\caption{Example path-integral geometry generating a BCFT state corresponding to a two-sided black hole system with dynamical energy density.}
\label{fig:SLQ}
\end{figure}

As a specific example, we can take the ``single joining quench'' geometry of \cite{Shimaji:2018czt} and add to it another boundary centered at the origin; this second boundary is taken to be the image of the inner boundary of the annulus under the conformal transformation
\begin{equation}
    w(\zeta) = \frac{2 \zeta}{1 - \zeta^{2}} \: ,
\end{equation}
which takes us from the unit disk (with complex coordinate $\zeta = x + i \tau$) to the single joining quench geometry (with coordinate $w = \hat{x} + i \hat{\tau}$). An example of the resulting path-integral geometry is shown in figure \ref{fig:SLQ}.

We note a few important features of such a map. Firstly, the symmetry $x \rightarrow -x$ translates to a symmetry $\hat{x} \rightarrow -\hat{x}$, and likewise symmetry $\tau \rightarrow -\tau$ translates to symmetry $\hat{\tau} \rightarrow -\hat{\tau}$. Secondly, the outer annular boundary $| \zeta | = 1$ maps to the intersection of the slits $i [1, \infty)$ and $-i [1, \infty)$, while the inner boundary maps to

\begin{equation}
    \hat{x}^{2} + \hat{\tau}^{2} = \frac{1}{2 \cosh^{2}(L) } \Big( 1 + \sqrt{ 1 + \frac{4 \hat{x}^{2}}{\tanh^{2}(L)}} \Big) \: .
\end{equation}
Finally, we note that the energy density with respect to Euclidean time $\hat{\tau}$ is defined by
\begin{equation}
    T(w) + \bar{T}(\bar{w}) = \frac{3}{4(1 + w^{2})^{2}} + \frac{3}{4(1 + \bar{w}^{2})^{2}} = \frac{3}{2} \Big( \frac{\hat{\tau}^{4} - 2(3 \hat{x}^{2} + 1) \hat{\tau}^{2} + (\hat{x}^{2} + 1)^{2}}{((1 + \hat{x}^{2} - \hat{\tau}^{2})^{2} + 4 \hat{x}^{2} \hat{\tau}^{2})^{2}} \Big) \: ;
\end{equation}
the Lorentzian analogue decays as we move away from the boundary which represents the black hole.

\begin{figure}
\begin{center}
\includegraphics[width=.8\textwidth]{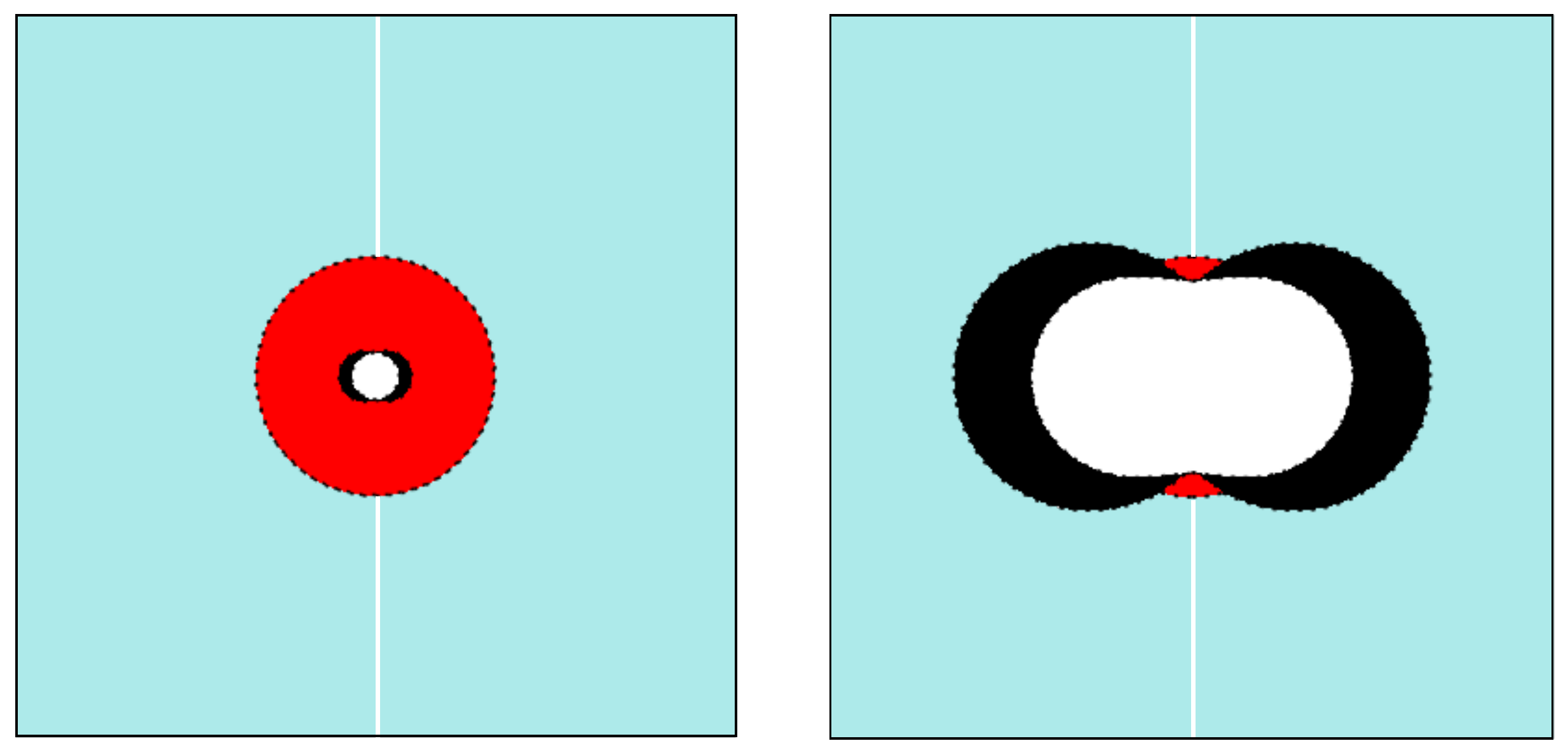}
\end{center}
\caption{Phase diagram for Euclidean modified (two boundary) single joining quench geometry with supercritical and subcritical $L$ respectively.}
\label{fig:euclideanslq}
\end{figure}

\begin{figure}
\begin{center}
\includegraphics[width=.8\textwidth]{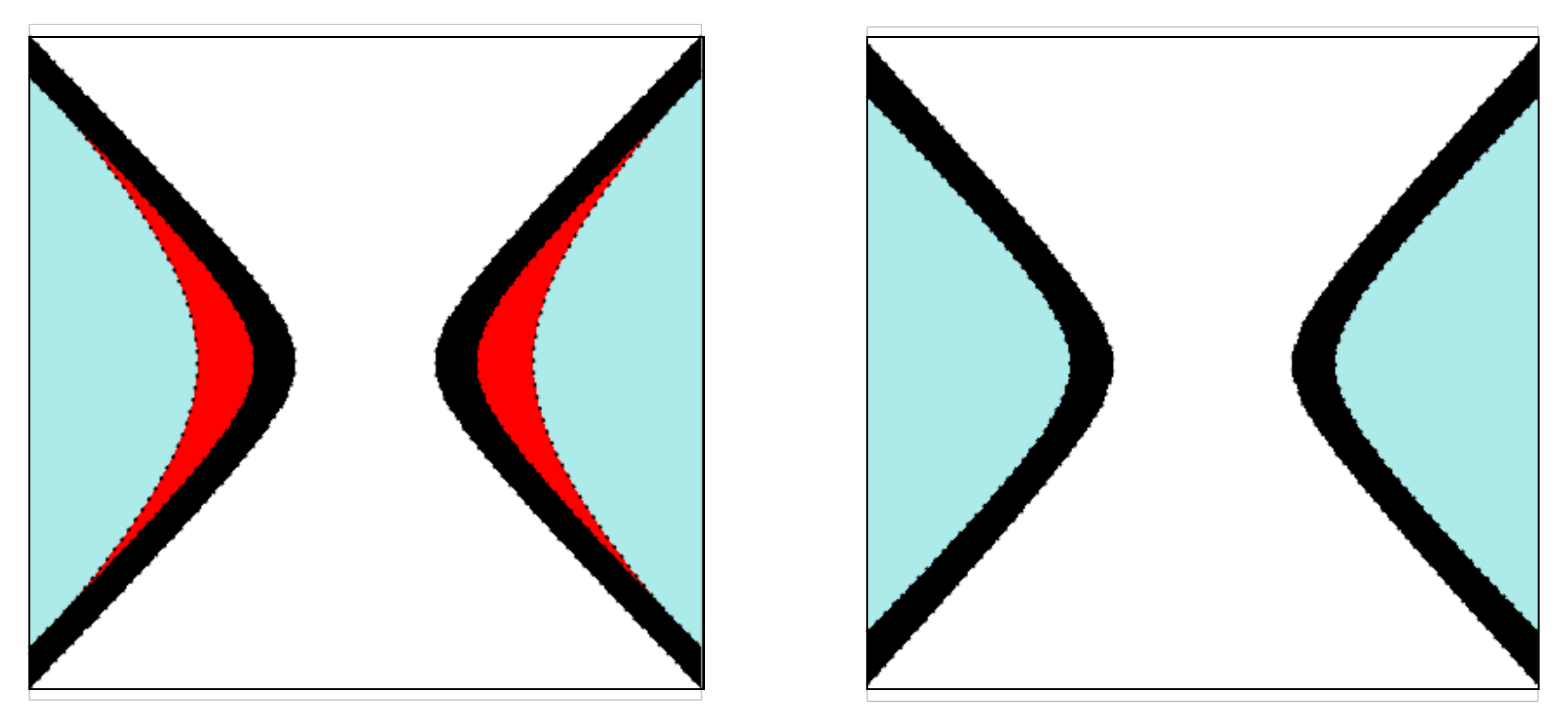}
\end{center}
\caption{Phase diagram for Lorentzian modified (two boundary) single joining quench geometry with supercritical and subcritical $L$ respectively.}
\label{fig:lorentzianslq}
\end{figure}

In the new coordinates, the phase boundary between connected HRT surfaces and disconnected surfaces ending on the outer ETW brane is $\hat{x}^{2} + \hat{\tau}^{2} = 1$, while the phase boundary between connected surfaces and disconnected surfaces ending on the inner ETW brane is
\begin{equation}
        \Big( \alpha (\hat{x}^{2} + \hat{\tau}^{2}) - \beta \hat{x} - \sin \Theta \Big)^{2} = (\hat{x}^{2} + \hat{\tau}^{2} + 1)^{2} - 4 \hat{\tau}^{2} \: ,
\end{equation}
with
\begin{equation}
\begin{split}
    \alpha & = \frac{(1+R^{2})^{2}(1 + \sin \Theta) - 4 R^{2}}{4 R^{2}} = \cosh^{2}(L) (1 + \sin \Theta) - 1 \\
    \beta & = \frac{(1+ R^{2})}{R} \cos \Theta = 2 \cosh(L) \cos \Theta \: .
\end{split}
\end{equation}
If a transition between the two disconnected phases is present, the phase boundary has locus
\begin{equation}
    \hat{x}^{2} + \hat{\tau}^{2} = \frac{2 \ell^{2}}{(1+\ell^{2})^{2}} \Big( 1 + \sqrt{ 1 + \frac{4 \hat{x}^{2} (1+\ell^{2})^{2}}{(1-\ell^{2})^{2}}} \Big)
\end{equation}
See Figure \ref{fig:euclideanslq}.
We can analytically continue $\hat{t} = - i \hat{\tau}$ to determine the BCFT boundaries and phase boundaries in Lorentzian signature. For $L > L_{\textnormal{c}}$, the phase boundaries now meet at the point
\begin{equation}
    \hat{x}_{0} = \frac{\alpha - \sin \Theta}{2 + \beta} \: , \qquad \hat{t}_{0} = \sqrt{\hat{x}_{0}^{2} - 1} \: .
\end{equation}
For $|\hat{t}| < \hat{t}_{0}$ we have three distinct phases, while for $|\hat{t}| > \hat{t}_{0}$ we just have the two disconnected phases. For $L < L_{\textnormal{c}}$, we just have the two disconnected phases (see Figure \ref{fig:lorentzianslq}).

One can now determine the time-dependence of the entanglement entropy
along any desired trajectory.
Recall from previous sections that, on the annulus, the HRT surfaces for
symmetrically situated intervals (with inner endpoints $(\pm x, \tau)$)
are circular arcs, and the corresponding entanglement entropy is given
by
\begin{equation}
S(x, \tau) = \begin{cases}
\ln \Big( \frac{2 x}{\tilde{\epsilon}(x, \tau)} \Big) \: , \qquad \qquad
\qquad \: \: \: \: \: \: \: \: \textnormal{connected}\\
\ln \Big( \frac{(x^{2} + \tau^{2}-R^{2})(1+\sin
	\Theta)}{\tilde{\epsilon}(x, \tau) R \cos \Theta} \Big) \: , \qquad
\textnormal{disconnected } T>0 \\
\ln \big( \frac{1 - x^{2} - \tau^{2}}{\tilde{\epsilon}(x, \tau)} \big)
\: , \qquad \qquad \qquad \: \: \: \textnormal{disconnected } T=0 \: ,
\end{cases}
\end{equation}
where we have recalled \cite{Caputa:2019avh} that the UV regulator
$\epsilon$ in the physical setup requires a position dependent regulator
$\tilde{\epsilon}(x, \tau) = |\zeta'(w)| \epsilon$ in the annular setup.
It is a simple matter to apply the appropriate conformal transformation
and Wick rotate to Lorentzian signature, whence we recover the
expression for the entanglement entropy of symmetrically situated
intervals in the Lorentzian modified local quench geometry.

\section{Discussion}

In this section we present a few additional observations and some directions for future work.

\subsection{A connection to behind-the-horizon physics of black hole microstates}

There is an interesting connection between the transitions in entanglement entropy that we have observed in this paper and another type of  transition for entanglement entropy pointed out in \cite{Cooper:2018cmb}. In that paper, the authors (including some of the present authors) considered black hole microstates for a holographic CFT on $S^d$ defined via a Euclidean path-integral on a finite cylinder, with a boundary at time $\tau_0$ in the Euclidean past. This corresponds to the evolution of a boundary state $|B\rangle$ by Euclidean time $\tau_0$. In the 2D CFT case for small enough $\tau_0$, this state corresponds to a single-sided black hole at temperature $4/\tau_0$, with a time-dependent ETW brane behind the horizon providing an inner boundary for the black hole.

For these states, the entanglement entropy for an interval can exhibit a phase transition as the interval size is increased, such that after the transition, the entanglement wedge of the interval includes a region behind the black hole horizon (terminating on the ETW brane). This is somewhat reminiscent of the entanglement wedge transition discussed in this paper, but it turns out that there is a precise connection between the two.

If we unwrap the circle on which the CFT lives, we obtain a planar black hole dual (above the Hawking-Page transition \cite{Miyaji:2014mca}) to the global quench geometry \cite{Calabrese:2016xau}. The holographic results for entanglement entropy in this situation are the same as in the compact case, since the gravity dual for the compact case is just a periodic identification of the gravity dual for the non-compact case.

The CFT calculation of entanglement entropy in the non-compact case is carried out via a correlation function of twist operators on an infinite strip. But a local conformal transformation maps this calculation to exactly the CFT calculation in section 3.2 used to deduce the phase transition in this paper.

\begin{figure}
\centering
\includegraphics[width=150mm]{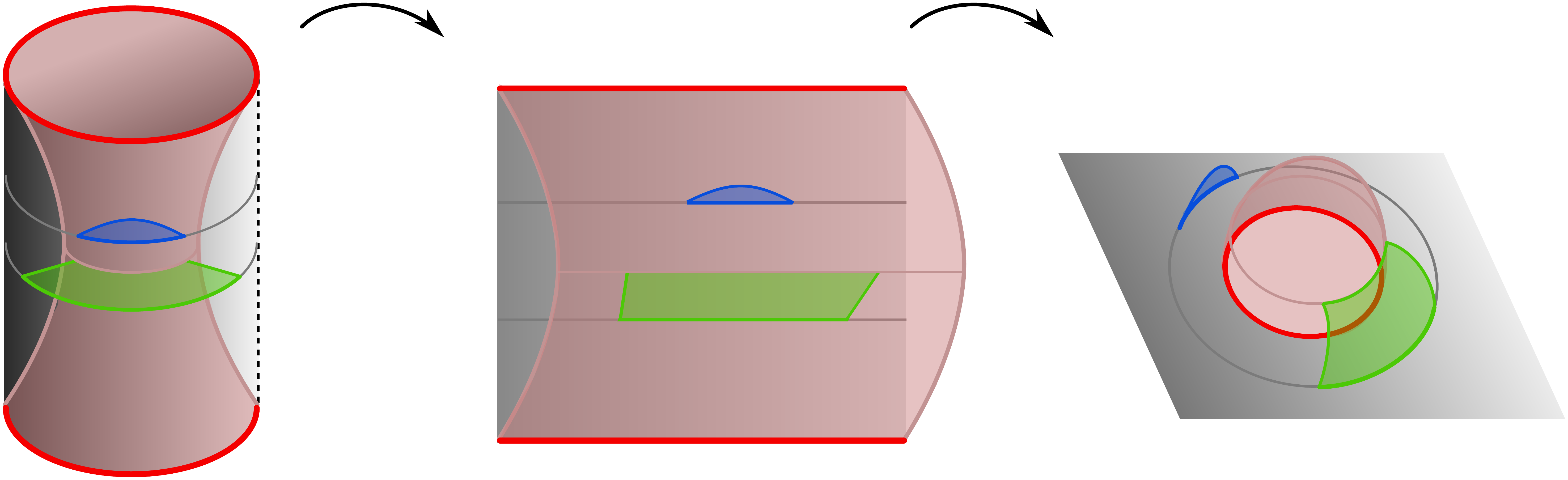}
\caption{BTZ black hole microstates have the same brane profile and hence
  entanglement entropy as the planar black hole dual
  to a global quench. The quench geometry is obtained from a local
  conformal transformation of the excised disk, so the transition in entanglement entropy for the static case described above, and the BTZ microstates in \cite{Cooper:2018cmb}, are controlled by the same CFT correlator.
}
\label{fig:microstate}
\end{figure}

We visual this connection in figure \ref{fig:microstate}. In the single-sided microstates, there is a transition in the extremal surfaces as the boundary region is increased (blue and green regions in figure
\ref{fig:microstate}).
In the CFT, this can be calculated by a correlator of twists in the large-$c$ limit and simple spectral constraints \cite{bcft2}.
Remarkably, this is essentially the same correlator governing the transition in entanglement wedge, as a function of subsystem size, as the static 2D case described in section 3.

\subsection{CFT constructions for duals of higher-dimensional evaporating black holes}

In future work, it would be interesting to study explicitly some higher-dimensional analogues of the constructions considered in this paper. We describe a few specific constructions in this final section. For these higher-dimensional examples, a detailed study will likely require some numerics as the bulk geometry will no longer be locally AdS. However, as the geometries depend on only two variables, such a study should be quite feasible.

\subsubsection*{BCFT microstate construction}

\begin{figure}
\begin{center}
\includegraphics[width=.8\textwidth]{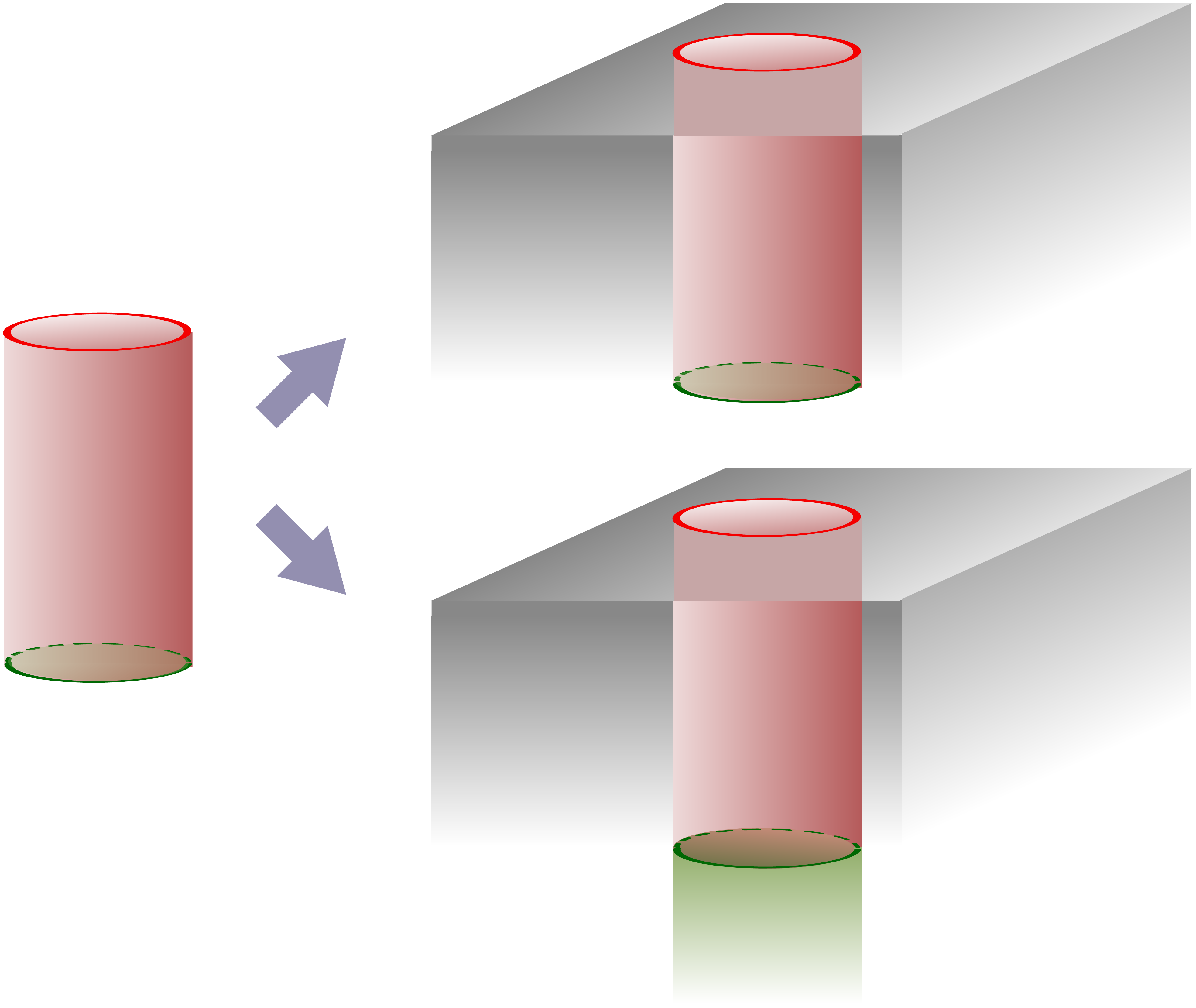}
\end{center}
\caption{Higher dimensional construction based on BCFT microstates}
\label{fig:BHMCJ}
\end{figure}

Figure \ref{fig:BHMCJ} shows on the left a Euclidean path integral for a high-energy CFT state obtained by placing some boundary conditions in the Euclidean past (at the red sphere). This corresponds to a black hole with some time-dependent behind-the-horizon physics, as described in \cite{Cooper:2018cmb}. We have in mind that the red boundary corresponds to a boundary condition with a large boundary entropy, so that the holographic description involves a brane with large tension.

Now we couple this system to a bulk CFT as shown on the right. Here, we need to introduce an additional boundary component (shown in green) into the Euclidean path integral. Two possible choices for the topology of this boundary component are shown. We have in mind that this boundary has a small boundary entropy, perhaps corresponding to a $T=0$ brane. This setup is the precise higher-dimensional analog of the single-sided setup of section 4.1.

In the dual holographic theory, using the bottom-up approach, we will have a bulk $d+1$-dimensional gravity action, but also two different types of $d$-dimensional ETW branes corresponding to the two different boundary conditions. Finally, there will be another $d-1$ dimensional brane that serves as the interface between the two types of $d$-dimensional branes. This can have its own tension parameter independent of the others.

\subsubsection*{Vaidya-type construction}

\begin{figure}
\begin{center}
\includegraphics[width=.5\textwidth]{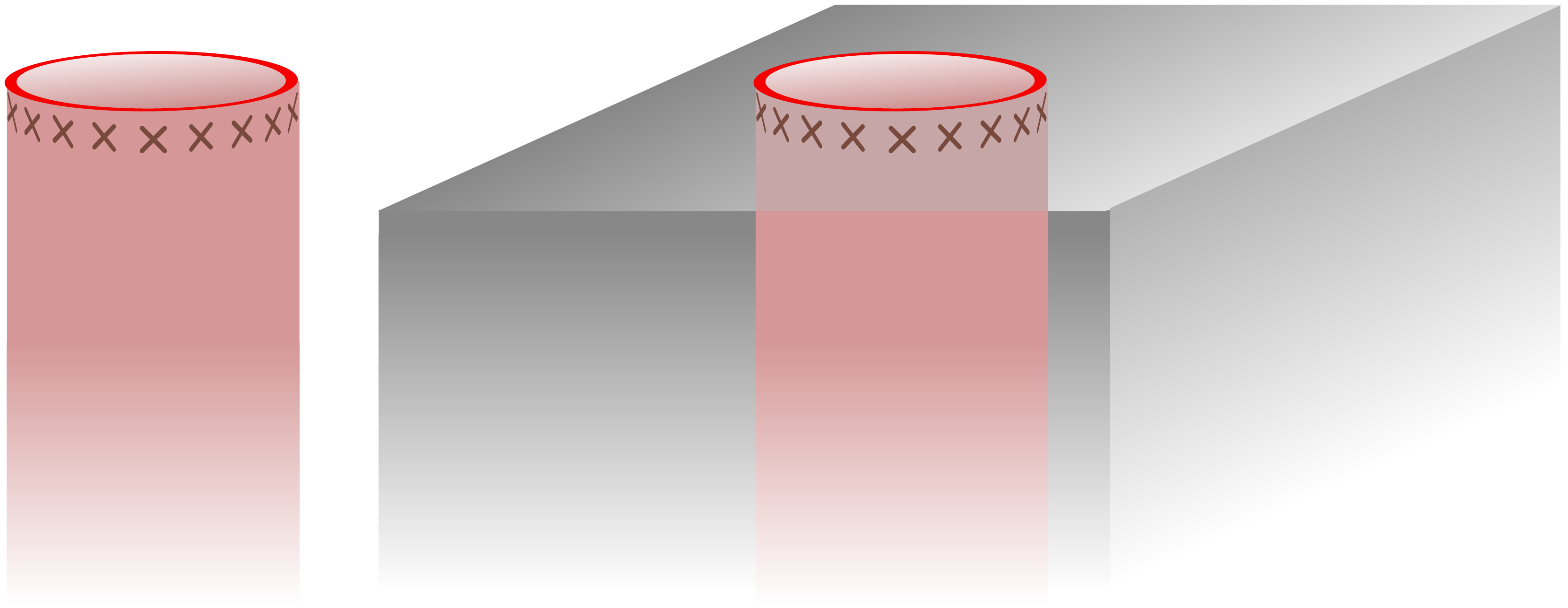}
\end{center}
\caption{Higher-dimensional construction based on CFT-Vaidya states.}
\label{fig:Hartman}
\end{figure}

Another interesting case makes use of the setup of \cite{Anous:2016kss}. Figure \ref{fig:Hartman} shows on the left a Euclidean path integral for a CFT state dual to a shell of matter that collapses to form a black hole. We have insertions of many operators at some small time in the Euclidean past. Alternatively, we could consider a smooth source for some operator, again localized around some particular time $\tau = -\epsilon$. We can take a limit where $\tau \to 0$ but the sources/insertions are chosen such that we end up with a finite energy state.

Now we couple this system to a bulk CFT as shown on the right. Without the sources, this path-integral would give the vacuum state of the BCFT. We expect that the sources mainly excite boundary degrees of freedom, so the bulk part of the CFT is still nearly in the vacuum state. In this case, we expect that the state is dual to a shell that collapses to form a black hole but then evaporates.

\section*{Acknowledgments}

We would like to thank Tarek Anous for pointing out that the Acknowledgments section was missing from the original version of the paper. We would like to thank Ahmed Almheiri, Jordan Cottler, Tom Hartman, Lampros Lamprou, and Jason Pollack for useful discussions. DW is supported by an International Doctoral Fellowship from the University of British Columbia.  MVR is supported by the Simons Foundation via the It From Qubit Collaboration and a Simons Investigator Award. This work is supported in part by the Natural Sciences and Engineering Research Council of Canada.

\bibliographystyle{jhep}
\bibliography{Evaporation}

\end{document}